\documentstyle[epsf,12pt]{article}

\font\boldgreek=cmmib10
\mathchardef\mymu="0916 
\def\R{ {\rm R \kern -.31cm I \kern .15cm}}
\def\C{ {\rm C \kern -.15cm \vrule width.5pt \kern .12cm}}
\def\Z{ {\rm Z \kern -.27cm \angle \kern .02cm}}
\def\N{ {\rm N \kern -.26cm \vrule width.4pt \kern .10cm}}
\def\1{{\rm 1\mskip-4.5mu l} }

\def \cc #1 {\kern .7em \hfill #1 \hfill \kern .7em}
\begin{document}
\begin{titlepage}
\parskip=0.2truecm
\centerline{\bf The Heavy Quark Potential from}
\medskip
 \centerline{\bf Wilson's Exact Renormalization Group}\bigskip

\bigskip
\centerline{{\bf Ulrich ELLWANGER}\footnote{e--mail:
ellwange@qcd.th.u--psud.fr}}
\centerline{Laboratoire de Physique Th\'eorique et Hautes
Energies\footnote{Laboratoire
associ\'e au Centre National de la Recherche Scientifique -- URA D0063}}
\centerline{Universit\'e de Paris XI, b\^atiment 211, 91405 Orsay Cedex,
France}
\bigskip

\centerline{{\bf Manfred HIRSCH}\footnote{Supported by a DAAD--fellowship HSP
II
financed by the German Federal Ministry for Research and Technology; e--mail:
m.hirsch@thphys.uni-heidelberg.de}
and {\bf Axel WEBER}\footnote{Supported by a
LGFG--fellowship of the Land Baden--W\"urttemberg
and a DAAD--fellow\-ship; current address: Instituto de Ciencias Nucleares, 
Universidad Nacional Au\-t\'o\-no\-ma de M\'exico, Circuito Exterior, 
A.P.\ 70--543, 04510 M\'exico, D.F., Mexico;
e--mail: axel@nuclecu.unam.mx
}}\par
\centerline{Institut  f\"ur Theoretische Physik}\par
\centerline{Universit\"at Heidelberg, Philosophenweg 16, D--69120 Heidelberg,
FRG}\par
\centerline{and}\par
\centerline{Laboratoire de Physique Th\'eorique et Hautes
Energies\footnotemark[2]} \par
\centerline{Universit\'e de Paris XI, b\^atiment 211, 91405 Orsay Cedex,
France} \par

\bigskip
\noindent
${\bf Abstract}$ \par
We perform a calculation of the full momentum dependence of the gluon and
ghost propagators in pure SU(3) Yang--Mills theory by integrating Wilson's 
exact renormalization group equations with respect to an infrared cutoff 
$k$. The heavy quark potential in the quenched approximation can be 
expressed in terms of these propagators. Our results strongly indicate a 
$1/p^4$--behaviour of the heavy quark potential for $p^2 \to 0$. We show
in general, that effective actions which satisfy Schwinger--Dyson equations,
correspond to (quasi--) fixed points of Wilson's exact renormalization
group equations.

\bigskip
\noindent LPTHE Orsay 96--50, HD--THEP--96--21, ICN--UNAM--96--08 \par

\end{titlepage}

\supereject
\noindent {\bf 1. Introduction}

\medskip
The computation of reliable phenomenological numbers on the basis of
non--perturba\-tive
QCD is still an unsolved problem. Numerical methods seem to be required,
as the
simulation of the theory on a space--time lattice. A
semi--phenomenological tool is
provided by the Schwinger--Dyson equations (SDEs), which are reviewed in
this respect in
[1]. In the present paper we discuss an approach to non--perturbative
phenomena in non--abelian gauge theories, which is based on the
integration of exact
renormalization group equations (ERGEs) [2] in continuum quantum field
theory [3]. \par

ERGEs describe the continuous evolution of effective Lagrangians, or effective
actions, with a scale (or infrared cutoff) $k$, and allow to obtain the
full quantum effective action for $k \to 0$ from a bare (microscopic)
effective action at a cutoff scale $k = \Lambda$. In contrast to
standard renormalization group
equations they describe this evolution including all irrelevant couplings, or
higher dimensional operators, and are exact in spite of the appearance of only
one--loop diagrams. The price to pay is the fact that one has to deal with an
infinite system of coupled differential equations which for practical
purposes
requires some truncation. The types of approximations or systematic
expansions,
which can be employed, depend on the phenomena under consideration: \par

One can expand the effective Lagrangian (or Hamiltonian or action) in
powers of momenta or derivatives, but keep all powers of the involved
field. This kind of expansion within the ERGEs is appropriate for the
calculation of effective potentials, which are required to study the
existence and nature of phase transitions (see [4] for some early
literature). \par

Alternatively, one can expand the effective action in powers of
fields, keeping all powers of the momenta [5]. This allows, e.g., to
study the formation of bound states by using ERGEs to look for the
appearance of poles in four--point functions [6].
Finally, in order to describe dynamical symmetry breaking,
composite
fields can be introduced [7, 8] and different expansions can be applied
to the parts
of the effective action involving fundamental and composite fields,
respectively. \par

The ansatz of the present paper is based on the expansion of the
effective action in powers of fields, which allows to study the full
momentum dependence of the (gluon and ghost) propagators. Within this
expansion the knowledge of the running 3-- and 4--point functions
is
required in order to integrate the ERGE for the running of the 2--point
function, the knowledge of the 4-- and 5--point functions is
required for
the ERGE of the 3--point function etc.\ [5, 6]. \par

In the case of gauge theories this expansion gets to some
extent reorganized: due to gauge invariance (or, more precisely, BRST
invariance) parts of higher N--point functions
(with certain Lorentz and gauge index structures)
are completely determined by lower N--point functions. 
The usual procedure consists in imposing Slavnov--Taylor identities 
(STIs) on the effective action, which can be solved for ``dependent''
parts of N--point functions in terms of ``independent'' ones. 
For instance, in the present case of a pure Yang--Mills theory a gauge
invariant term in the effective action of the form
$$ F_{\mu \nu}^a (f(D^2) F_{\mu \nu})^a \eqno(1.1) $$
(where $D$ denotes the covariant derivative and $f$ an arbitrary
function) describes simultaneously the ``independent'' effective gluon 
propagator or 2--point function, and certain ``dependent'' Lorentz and 
gauge index structures of the 3--, 4-- and higher gluonic N--point
functions. Additional ``independent'' structures of the gluonic 3-- and 
higher N--point
functions are described in terms of gauge invariant expressions
involving at least three powers of the field strength $F_{\mu \nu}$. 

The application of ERGEs to gauge theories faces some technical problems,
since ERGEs are based on an intermediate infrared momentum cutoff $k$,
and such a cutoff generally breaks gauge or BRST invariance [9--14]. One 
way to deal with these problems is the modification of the STIs [11, 13, 14]: 
the modified STIs impose
``fine tuning conditions'' on those couplings in the effective action at
$k \neq 0$, which break gauge or BRST invariance. These ``fine tuning
conditions'' are such that they guarantee the BRST invariance of the
effective action for $k \to 0$. 

As a consequence of the (modified) STIs, at any value of $k$, 
the ``dependent'' parts of the effective action can thus be obtained 
either via the modified STIs in terms of the 
``independent'' parts at the scale $k$, or directly from the integration of 
the ERGEs. Below we will employ approximations which consist in neglecting
 contributions to the r.h.sides of the ERGEs. In the absence of 
approximations, both methods to obtain the dependent parts lead to identical 
results. In the presence of approximations one should determine the dependent 
parts of the effective action from the modified STIs. Approximations 
on the r.h.sides of ERGEs for independent parts of the effective action 
do thus not imply a violation of the modified STIs. 

In the presence of approximations one can nevertheless study the ERGEs for 
the dependent 
parts of the effective action and see, whether these ERGEs deviate strongly 
from the corresponding equations obtained from the $k$--derivatives of the 
modified STIs. As emphasized in [14], this method allows for a numerical
 self--consistency check of 
the approximation and will also be employed in the present paper.

As described in section 3, our approximation consists in the neglect of
the contribution of certain parts of the 3-- and 4--point functions to 
the ERGEs of the 2--point functions. We take those parts of the 3-- and 
4--point functions into account, however, which are determined in terms 
of the 2--point functions through the standard part of the STIs.
In the purely gluonic sector these are those obtained by an
expansion of eq. (1.1) in powers of the gauge field $A_{\mu}^a$.

No approximation is imposed on the momentum dependence of the 2--point 
functions (the gluon and ghost propagators).  
These propagators by themselves are gauge dependent and not directly
observable quantities. A certain combination thereof, however, has a
physical meaning: The potential between infinitely heavy quarks can be
written as a product of the dressed quark--gluon vertices and the 
full gluon propagator, and the quark--gluon vertex function is given by 
the ghost propagator function. As a result the heavy quark potential 
depends on both propagator functions, which we calculate here in the
quenched approximation.

The aim of this paper is to see, how much information on the heavy quark 
potential in quenched QCD can be obtained within the approximation we 
employ, which is seen to be self--consistent for a range of 
momenta $p^2$ down to a certain value $p^2 \sim {k_0}^2$ (see section 5). 
We find strong evidence for a $1/p^4$--behaviour within the trustworthy 
range of $p^2$. \par

Both the method and the results can be compared with the Schwinger--Dyson
formalism. To this end we derive in section 2 a new and general relation
between ERGEs and SDEs: we show that effective actions, which satisfy
SDEs in the presence of an infrared cutoff $k$, are (quasi--) fixed points
of the ERGEs. Thus a search for fixed points of the ERGEs for $k \to 0$ is
equivalent to the search for solutions of SDEs. \par

Investigations of the gluon propagator have been performed in the
context of SDEs both in the axial [15, 16] and Landau [17] gauge. In
most cases solutions with a singular behaviour like $1/p^4$ have been
found to be consistent. (A behaviour less singular than $1/p^2$ has been
put forward on phenomenological grounds in [18] and claimed to exist in
the axial gauge in [16], but recently such a form of the gluon
propagator in the infrared has been argued to be inconsistent [19].) In
the Landau gauge the ghost sector has been completely neglected; thus it
is much
less straightforward to relate these results to physical observables. In
section 6 we will discuss the comparison between the ERGE and SDE
approach in more detail. \par

The paper is organized as follows: in section 2 we discuss the
derivation of the
ERGEs, and their relation to SDEs.
In section~3 we present our truncated
ansatz for the pure Yang--Mills action. In section~4 we discuss
the technical procedure, and in section 5 our
numerical results. Section~6 is devoted to conclusions and an outlook.

\par \bigskip
\noindent {\bf 2. ERGEs and SDEs}

\medskip
The form of the ERGEs can most conveniently be derived assuming a path integral
representation for the generating functional of the theory. To this end
we restrict
ourselves, for the time being, to the case of a single scalar field. It is
straightforward to supplement the following equations with the
appropriate minus signs,
Lorentz and internal indices in the cases of fermions, vector fields and
internal
symmetries. To start with, the path integral formulation of the theory
requires for its proper
definition a regularization of the ultraviolet divergences. This will
be implemented
through a modification of the propagator, i.e.\ the term
quadratic in the fields in the bare action.
In addition we introduce an infrared cutoff $k$; the ERGEs will describe
the variation
of the generating functional with $k$, where the UV cutoff $\Lambda$
will be held
fixed. (In the framework of the ERGEs, in particular in the application
to gauge
theories, one can consistently set $\Lambda = \infty$, see below).
The UV and IR
regularized generating functional of connected Green functions ${\cal
G}_k^{\Lambda}(J)$ can then
be represented as
$$e^{-{\cal G}_k^{\Lambda}(J)} = {\cal N} \int {\cal D} \varphi \ e^{-
{1 \over 2}
(\varphi , P_0^{-1} \varphi ) - \Delta S_k^{\Lambda} - S_{int}(\varphi )
+ (J, \varphi )}
\ \ \ , \eqno(2.1)$$
where $S_{int}(\varphi)$ is independent of $k$.
Here $(J, \varphi )$ etc. is a short--hand notation for
$$(J, \varphi ) \equiv \int {d^4p \over (2 \pi )^4} \, J(p) \varphi (-p)
\ \ \ . \eqno(2.2)$$
\noindent $P_0^{-1}$ denotes the bare inverse propagator (e.g.
$P_0^{-1}(p^2) = p^2$ in
the case of a massless scalar field), $S_{int}$ the bare interaction,
and
$$\Delta S_k^{\Lambda} = {1 \over 2} (\varphi , R_k^{\Lambda} \ \varphi )
\eqno(2.3)$$
implements the UV and
IR cutoffs. In the case of a massless field a convenient choice for
$R_k^{\Lambda}$ is
$$R_k^{\Lambda}(p^2) = p^2 \, {1 - e^{-{p^2 \over \Lambda^2}} + e^{- {p^2
\over k^2}} \over
e^{- {p^2 \over \Lambda^2}} - e^{- {p^2 \over k^2}}} \ \ \ , \eqno(2.4)$$
\noindent such that the full propagator
$$P_k^{\Lambda} = \left ( P_0^{-1} + R_k^{\Lambda} \right )^{-1} =
{e^{-{p^2 \over
\Lambda^2}} - e^{-{p^2 \over k^2}} \over p^2} \eqno(2.5)$$
\noindent has the desired properties: exponential decay in the UV, and
finiteness in the IR. \par

The ERGE for ${\cal G}_k^{\Lambda}$ describes its variation with $k$, at
fixed $\Lambda$.
It can be obtained by differentiating eq.\ (2.1) with respect to $k$, and
replacing
$\varphi$ under the path integral by variations with respect to the
sources. One finds
$$
\partial_k {\cal G}_k^{\Lambda}(J) = - {1 \over 2} \int {d^4q \over (2
\pi )^4} \,
\partial_k R_k^{\Lambda}(q^2) \cdot \left \{ {\delta^2 {\cal
G}_k^{\Lambda}(J) \over
\delta J(q) \ \delta J(-q)} - {\delta {\cal G}_k^{\Lambda}(J) \over
\delta J(q)} {\delta
{\cal G}_k^{\Lambda}(J) \over \delta J(-q)} \right \} \ \ \ . \eqno(2.6)
 $$
\par
\noindent The effective action $\Gamma_k^{\Lambda}(\varphi )$ is defined
by the Legendre
transform of ${\cal G}_k^{\Lambda}(J)$,
$$\Gamma_k^{\Lambda}(\varphi ) = {\cal G}_k^{\Lambda}(J) + (J, \varphi )
\ \ \ ,
\eqno(2.7)$$
and it is convenient to define an effective action 
$\widehat{\Gamma}_k^\Lambda$ with the IR cutoff terms subtracted:
$$\widehat{\Gamma}_k^{\Lambda} = \Gamma_k^{\Lambda} - \Delta S_k^{\Lambda}
\eqno(2.8) $$
\noindent Inserting the Legendre transformation into eq.\ (2.6) one
obtains the ERGE for
$\widehat{\Gamma}_k^{\Lambda}$:
$$\partial_k \widehat{\Gamma}_k^{\Lambda}(\varphi ) = {1 \over 2} \int {d^4q
\over (2 \pi )^4} \,
\partial_k R_k^{\Lambda}(q^2) \cdot
\left (
{\delta^2 \widehat{\Gamma}_k^{\Lambda}(\varphi ) \over \delta \varphi (q) \delta
\varphi (-q)} + R_k^{\Lambda}(q^2)
\right )^{-1} \ \ \ . \eqno(2.9)$$
\par
\noindent The boundary condition of $\widehat{\Gamma}_k^{\Lambda}(\varphi )$
for $k \to \Lambda$ can be obtained from a careful consideration of the
limit $P_k^{\Lambda} \to 0$ for $k \to \Lambda$ and reads
$$\widehat{\Gamma}_{\Lambda}^{\Lambda}(\varphi ) = {1 \over 2} \left(
\varphi, ( P_0^{\Lambda} )^{-1}
\varphi \right) +
S_{int}(\varphi ) \ \ \ . \eqno(2.10)$$
\noindent On the other hand, for $k \to 0$ $\widehat{\Gamma}_k^{\Lambda}
(\varphi)$ becomes the
full quantum effective action of a theory, which is defined by a bare
interaction
$S_{int}(\varphi )$ and a fixed UV cutoff $\Lambda$. Since the
integration of the ERGEs
allows us to relate effective actions $\widehat{\Gamma}_k^{\Lambda}$ with
different $k$, it can be
used to compute $\widehat{\Gamma}_0^{\Lambda}$ in terms of
$\widehat{\Gamma}_{\Lambda}^{\Lambda}$ related to
$S_{int}$ by (2.10). \par

For most applications in particle physics this is actually all we want:
there always exists
a scale $\Lambda$ (at least the Planck scale) beyond which we
expect unknown new
interactions or particles with masses larger than $\Lambda$. Physics
beyond $\Lambda$,
including quantum fluctuations involving momenta $p^2$ with $p^2 \geq
\Lambda^2$, will
certainly affect the form of $S_{int}$ or the effective action
$\widehat{\Gamma}_{\Lambda}^{\Lambda}$. Given an ansatz for
$\widehat{\Gamma}_{\Lambda}^{\Lambda}$, however,
we are generally interested in the effect of quantum fluctuations
involving momenta with
$0 \leq p^2 \leq \Lambda^2$. If we integrate the ERGE for
$\widehat{\Gamma}_k^{\Lambda}$ from $k =
\Lambda$ down to $k = 0$, these quantum fluctuations have been fully
taken into account.
\par

Next we turn to the relation between ERGEs and SDEs. To this end it is
first convenient
to write the ERGEs in terms of the generating functional (or partition
function)
$Z_k^{\Lambda}(J)$, where $Z_k^{\Lambda}(J)$ is given by
$$Z_k^{\Lambda}(J) = e^{-{\cal G}_k^{\Lambda}(J)} \eqno(2.11)$$
\noindent and the r.h.side of eq.\ (2.11) by the path integral
(2.1). The ERGE for
$Z_k^{\Lambda}(J)$ reads
$$\partial_k Z_k^{\Lambda}(J) = - {1 \over 2} \int {d^4q \over (2 \pi
)^4} \, \partial_k
R_k^{\Lambda}(q^2) {\delta^2 Z_k^{\Lambda}(J) \over \delta J(q) \delta
J(-q)} \ \ \ .
\eqno(2.12)$$

The full set of SDEs can also most conveniently be expressed in terms of
$Z(J)$ [20].
They follow from the vanishing of the path integral over a total
derivative with respect to
$\varphi$. Subsequently we will consider the SDEs in the presence of an
UV cutoff
$\Lambda$ and an IR cutoff $k$ as in (2.1). In terms of
$Z_k^{\Lambda}(J)$ they read
$$ \left ( \left . J(p) - \left ( P_0^{-1}(p^2) + R_k^{\Lambda}(p^2)
\right ) {\delta \over
\delta J(-p)} - {\delta S_{int}(\varphi ) \over \delta \varphi (-p)}
\right |_{\textstyle \varphi =
{\delta \over \delta J}} \right ) Z_k^{\Lambda}(J) = 0 \ \ \ .
\eqno(2.13)$$
\noindent After replacing $Z_k^{\Lambda}$ by ${\cal G}_k^{\Lambda}$
according to eq.\
(2.11), and ${\cal G}_k^{\Lambda}$ by $\Gamma_k^{\Lambda}$ according to
eq.\ (2.7), one
obtains the SDEs for the one--particle irreducible vertices, which are
generally
considered in the literature for $k = 0$ and $\Lambda = \infty$. 

Let us now denote the entire l.h.side of eq.\ (2.13)
by $\Sigma_k^{\Lambda}(J)$. Next we evaluate the $k$--derivative of
$\Sigma_k^{\Lambda}$.
The $k$--derivative hits $R_k^{\Lambda}$ and $Z_k^{\Lambda}$, and we have
to use the ERGE
(2.12). The result can be written in the form
$$\partial_k \Sigma_k^{\Lambda} = - {1 \over 2} \int {d^4q \over (2 \pi
)^4} \, \partial_k
R_k^{\Lambda}(q^2) {\delta^2 \Sigma_k^{\Lambda} \over \delta J(q) \delta
J(-q)}  \ \ \ .
\eqno(2.14)$$
\noindent Thus we can make the following important statement: suppose
that we managed
to find a functional $Z_k^{\Lambda}(J)$, which satisfies the SDEs (2.13)
for some value
of $k$, identically in the sources $J$. Thus the corresponding quantity
$\Sigma_k^{\Lambda}(J)$ vanishes. Now let us start to vary $k$. Since
the r.h.side of eq.\ (2.14) vanishes, $\Sigma_{k'}^{\Lambda}$ will still 
vanish for $k' \not= k$.
(Here, of course, it has been used that $Z_k^{\Lambda}$ varies with $k$
according to the
ERGE (2.12).) Thus we find that $Z_{k'}^{\Lambda}$ still satisfies the
SDE (2.13), with
$k$ replaced by $k'$. In other words, if we found a functional
$Z_k^{\Lambda}$ which
satisfies the SDE (2.13) for some value of $k$, and compute
$Z_{k'}^{\Lambda}$ by
integrating the ERGE (2.12), $Z_{k'}^{\Lambda}$ is guaranteed to satisfy
the SDEs for all
values for $k'$. The same statement will hold for the other generating
functionals ${\cal
G}_k^{\Lambda}$ and $\Gamma_k^{\Lambda}$. Thus generating functionals
which satisfy
$k$--dependent SDEs can be named quasi--fixed points of the ERGEs
(``quasi'' because they still depend on $k$). In particular, for $k \to
0$, ${\cal G}_0^{\Lambda}$ and $\Gamma_0^{\Lambda}$ will satisfy the
standard
SDEs (without an infrared cutoff, but still an ultraviolet cutoff
$\Lambda$) if such a quasi--fixed point is approached in the infrared.
\par

Note that this property is independent of the form of $S_{int}$ in eq.\
(2.13): Any SDE with any form of $S_{int}$ in eq.\ (2.13) plays the
role of a quasi--fixed point. (In particular, since we work in the
presence of an UV cutoff $\Lambda$, $S_{int}$ can contain arbitrary
higher dimensional interactions suppressed by powers of $\Lambda$.)
\par

These formal arguments let us suspect, that results obtained by the
integration of the ERGEs can be quite similar to results obtained by 
solving SDEs, although the formal arguments no longer hold in  a strict 
sense as soon as the effective actions are approximated or truncated in 
some way. \par

Let us note, for later use, that on the r.h.sides of the ERGEs
(2.6) and (2.9) we
can replace $R_k^{\Lambda}(p^2)$ by $R_k(p^2)$,
where $R_k(p^2)$
is defined by
$$R_k(p^2) = R_k^{\Lambda = \infty}(p^2) \ \ \ . \eqno(2.15)$$
\noindent In the limit $\Lambda \to \infty$ the momentum
integrations on the r.h.sides of (2.6) and (2.9) are still
exponentially damped with the choice (2.4) for $R_k^{\Lambda}$.
Accordingly we denote by ${\cal G}_k$ and $\widehat{\Gamma}_k$ the 
generating functionals, which
satisfy the ERGEs with $R_k^{\Lambda}$ replaced by $R_k$. Of course we
are still able to
compute $\widehat{\Gamma}_{k=0}$ in terms of an ansatz for 
$\widehat{\Gamma}_{k=\Lambda}$.
In physical terms UV divergences do not appear, since 
$\widehat{\Gamma}_{\Lambda}$ is already supposed to contain all
quantum effects
involving momenta with $p^2 \geq \Lambda^2$. We just loose the
possibility to formally equate $\widehat{\Gamma}_{\Lambda}$ with 
$S_{int}$ appearing in a path integral of
the form of eq.\
(2.1), although a slightly different, more involved path integral
representation for $\widehat{\Gamma}_k$ still exists [21]. From 
universality we expect in
any case, that $\widehat{\Gamma}_0$ is independent of minor modifications of
$\widehat{\Gamma}_{\Lambda}$, so
one choice is a priori as good as the other. Thus, subsequently, we
define a theory by
the ERGEs (2.6) or (2.9), with $R_k^{\Lambda}$ replaced by $R_k$, and a
finite ansatz for
$\widehat{\Gamma}_{\Lambda}$ (or ${\cal G}_{\Lambda}$) at a fixed scale
$\Lambda$.\par \bigskip

\noindent {\bf 3. Yang--Mills Theories}

\medskip
The application of ERGEs to gauge theories requires some treatment of the
breaking of
gauge invariance, which is introduced through the intermediate IR cutoff
$k$. Here we
use the method of imposing modified STIs [11, 13,
14] in the
formulation presented in [13, 14]. In ref.\ [14], in particular, ERGEs for
SU(3) Yang--Mills theory have already been integrated numerically within
an approximation,
where only marginal and relevant couplings in the effective action have
been taken into
account. \par

In the present paper we extend the form of the effective
action considerably,
namely we allow for an arbitrary momentum dependence of the gluon and ghost
pro\-pa\-gators. Again we consider the SU(3) Yang--Mills theory in
four--dimensional
Euclidean space--time, where the classical action reads

$$
S  =  \int\mbox{d}^4
x\left\{{\textstyle\frac{1}{4}}F^a_{\mu\nu}
   F^a_{\mu\nu} + \frac{1}{2\alpha}\partial_\mu A^a_\mu\partial_\nu A^a_\nu
   + \partial_\mu\bar{c}^a\left(\partial_\mu c^a + g f^a{}_{bc}A^b_\mu c^c
   \right)\right .\nonumber $$
$$  \left .
    - K^a_\mu\left(\partial_\mu c^a + g f^a{}_{bc}A^b_\mu c^c\right) - L^a
   {\textstyle\frac{1}{2}}g f^a{}_{bc}c^b c^c +
\bar{L}^a\frac{1}{\alpha}\partial_\mu
   A^a_\mu \right\}
\eqno(3.1)
$$

with
$$
  F^a_{\mu\nu} = \partial_\mu A^a_\nu - \partial_\nu A^a_\mu
   + g f^a{}_{bc}A^b_\mu A^c_\nu \:.
\eqno(3.2)
$$
We have included the usual gauge fixing and ghost parts and coupled
external
sources to the BRST variations

$$
\delta A^a_\mu  =  \left(\partial_\mu c^a + g f^a{}_{bc}
   A^b_\mu c^c\right)\zeta $$
$$ \delta c^a  =  {\textstyle\frac{1}{2}}g f^a{}_{bc}c^b c^c\zeta
$$
$$
   \delta \bar{c}^a  =  -\frac{1}{\alpha}\partial_\mu A^a_\mu\zeta \:,
\eqno(3.3)
$$

\noindent where $\zeta$ is a Grassmann parameter. The invariance of $S$ at 
$\bar{L} = 0$ under BRST transformations can then be expressed as

$$
0  =  \left.\int\mbox{d}^4 x\left\{\delta
A^a_\mu\frac{\delta S}
   {\delta A^a_\mu} + \delta c^a\frac{\delta S}{\delta c^a} + \delta\bar{c}^a
   \frac{\delta S}{\delta\bar{c}^a}\right\}\right|_{\bar{L} = 0}$$
$$
    =  \zeta\left.\int\mbox{d}^4 x\left\{\frac{\delta S}{\delta K^a_\mu}
   \frac{\delta S}{\delta A^a_\mu} - \frac{\delta S}{\delta L^a}
   \frac{\delta S}{\delta c^a} - \frac{\delta S}{\delta\bar{L}^a}
   \frac{\delta S}{\delta\bar{c}^a}\right\}\right|_{\bar{L} = 0} \:.
\eqno(3.4)
$$

IR cutoffs in the gluon and ghost propagators will be introduced by
adding a term of
the form
$$\Delta S_k = {1 \over 2} ( A_{\mu}^a, R_{k, \mu \nu} \ A_{\nu}^a
) + ( \bar{c}^a , \tilde{R}_k \ c^a ) \eqno(3.5)$$
\noindent to the action. Explicit expressions for $R_{k, \mu \nu}$ and
$\tilde{R}_k$
will be given below. The effective action
$$\Gamma_k(A, c, \bar{c}, K, L, \bar{L}) \equiv \widehat{\Gamma}_k +
\Delta S_k
\eqno(3.6)$$
\noindent satisfies an ERGE of the form [13, 14]
$$
\partial_k \widehat{\Gamma}_k = \int {d^4p \over (2 \pi )^4} \left \{ {1
\over 2}
\partial_k R_{k,\mu \nu}(p^2)
\left ( \Gamma_k^{(2)}\right )^{-1}_{A_{\nu}^a (-p), A_{\mu}^a(p)}
\right .$$
$$\left .
{}- \partial_k \tilde{R}_k(p^2) \left ( \Gamma_k^{(2)}\right )^{-1}_{
-\bar{c}^a(-p), c^a (p)}
\right\} \ .
\eqno(3.7)
$$
\noindent Here $\left( \Gamma_k^{(2)} \right)^{-1}_{A^a_\mu (-p), A^b_\nu
(q)}$ is the $(A^a_\mu (-p), A^b_\nu (q))$--component of the inverse of
the matrix $\Gamma_k^{(2)} = \delta^2 \Gamma_k/\delta \bar{\varphi} \, \delta
\varphi$ of second derivatives of $\Gamma_k$ with respect to the fields
$\bar{\varphi}^B = \{A^a_\mu(-p), -\bar{c}^a(-p), c^a(-p)\}$ and $\varphi^B 
= \{A^a_\mu(p), c^a(p), \bar{c}^a(p)\}$, where the index of
$\bar{\varphi}$ and $\varphi$ runs over the different fields, momenta,
and Lorentz and gauge indices.
For $k \to 0$, $\widehat{\Gamma}_0$ has to satisfy the STI (3.4). To this end
$\widehat{\Gamma}_k$ has to satisfy, in particular at the starting point
$k = \Lambda$, a
modified STI of the form [13, 14]
$$\left. \int {d^4p \over (2 \pi )^4} \left \{ {\delta \widehat{\Gamma}_k 
\over \delta K_{\mu}^a(-p)} \ {\delta \widehat{\Gamma}_k \over \delta
A_{\mu}^a (p)} -
{\delta \widehat{\Gamma}_k \over \delta L^a (- p)} \ {\delta
\widehat{\Gamma}_k \over \delta c^a(p)} - {\delta \widehat{\Gamma}_k \over 
\delta \bar{L}^a (- p)} \ {\delta \widehat{\Gamma}_k \over \delta 
\bar{c}^a(p)} \right \} \right|_{\bar{L} = 0} $$
$$ = \int {d^4p \over (2 \pi )^4} \sum_{B} \left [ R_{k , \mu \nu}(p^2) 
{\delta^2 \widehat{\Gamma}_k
\over \delta K_{\nu}^a (-p) \ \delta {\varphi}^B } \left ( \Gamma_k^{(2)}
\right )^{-1}_{\bar\varphi^B, A_{\mu}^a(p)} \right .$$
$${}- \tilde{R}_k(p^2) {\delta^2 \widehat{\Gamma}_k \over \delta L^a(-p) \
\delta {\varphi}^B} \left ( \Gamma_k^{(2)}\right )^{-1}_{\bar\varphi^B,
c^a(p)} $$
$$ \left.\left. {}- \tilde{R}_k(p^2) {\delta^2 \widehat{\Gamma}_k
\over \delta \bar{L}^a(-p) \ \delta {\varphi}^B} \left ( \Gamma_k^{(2)}
\right )^{-1}_{\bar\varphi^B, \bar c^a(p)} \right ] 
\right|_{\bar{L} = 0} \ \ \ . \eqno(3.8)$$
\noindent The crucial point is that once eq.\ (3.8) is satisfied for $k =
\Lambda$, it
will be satisfied by $\widehat{\Gamma}_k$ for any $k < \Lambda$ provided
$\widehat{\Gamma}_k$ is obtained
from $\widehat{\Gamma}_{\Lambda}$ by integrating the ERGE. In particular
$\widehat{\Gamma}_{k=0}$ will
satisfy the standard STI (3.4), if the IR cutoff functions
$R_{k, \mu \nu}$ and
$\tilde{R}_k$ vanish identically for $k \to 0$. \par

Most importantly, $\widehat{\Gamma}_k$ at the starting point $k =
\Lambda$ can thus not be identified with the classical action (3.1), but
it has to contain symmetry breaking terms specified by the need to
satisfy eq.\ (3.8). The origin of these symmetry breaking terms can be
understood from the interpretation of $\widehat{\Gamma}_{\Lambda}$ as a
quantum effective action, where momenta $p^2$ with $p^2 \geq \Lambda^2$
have already been integrated out; the symmetry breaking induced by this
``IR'' cutoff $\Lambda$ then generates the symmetry breaking terms in
$\widehat{\Gamma}_{\Lambda}$. \par 

As stated above, our approach towards the integration of the ERGE (3.7)
is based on an expansion in powers of the gluon and ghost fields. We are
interested in the form of the gluon and ghost propagators, 
since a certain combination thereof (see below)
determines the heavy quark potential in the quenched approximation. 
It is thus essential to allow for an arbitrary momentum dependence of
the 2--point functions in our ansatz for the effective action.

Expanding the ERGE (3.7) in powers of fields one easily finds that
the full 3--gluon, 4--gluon, ghost--gluon and 4--ghost vertices appear
on the r.h.sides of the ERGEs for the gluon and ghost
propagators. In the sequel we will refer to parts of the vertices with
certain Lorentz and gauge index structures as ``operators''. One can
distinguish two different kinds of such operators:
\begin{enumerate}
\item[a)] ``Independent'' operators, which are not fixed in terms of the 
propagators or other operators by the STI. An example is
the contribution to the 4--gluon vertex obtained from
a term involving four powers of the field strength $F_{\mu\nu}^a$
in the effective action. (The r.h.side of the corresponding ERGE depends
on the full 2--, 3--, 4--, 5-- and 6--point functions.)
\item[b)] ``Dependent'' operators, which are fixed by the modified STI in 
terms of the propagators and operators of type a). In general eq.\ (3.8) 
yields a system of coupled non--linear integral equations for the dependent 
operators. Examples are 
the operators corresponding to those parts of the 3-- and 4--gluon
vertices, which are obtained by an expansion of the expression
$F_{\mu\nu}^a(f(D^2)F_{\mu\nu})^a$ in powers of the gauge field
(see eq.\ (1.1)), and a gluonic mass term (which we do not consider as
part of the gluon propagator in this context).
\end{enumerate}

Since our goal is the determination of the propagators, the only important
operators for our purpose are those, that contribute significantly to the
ERGEs for the 2--point functions. In a first approximation, we will
neglect the contributions of the operators of type a), which are the ones 
not determined by the
STI. Such an approximation is only expected to hold for a limited range
of scales $k$, and can to some extent be checked a posteriori (see
below). In any case, it turns the ERGE (3.7) into a closed system of
integro--differential equations for the propagators, which can be
integrated numerically.

Furthermore, with the simpler ansatz used in ref.\ [14] it has proved to
be permissible in a quantitative sense to determine the operators of
type b) through the standard form of the STI (eq.\ (3.8) with a 
vanishing r.h.side) instead of using the modified STI, with the important
exception of the gluon mass term. Again, this may only be true within a
certain range of $k$. We will therefore proceed in the same way here, i.e.\
we assume that we can approximate the operators of type b) by the
corresponding ones obtained from the standard STI, as far as their
quantitative contribution to the flow of the 2--point functions is
concerned.

In practical terms, this assumption implies a tremendous technical
simplification in the determination of these operators. For example,
for the 3-- and 4--gluon vertices we can use the expressions resulting
from an expansion of $F_{\mu\nu}^a(f(D^2)F_{\mu\nu})^a$ in powers of
$A^a_\mu$ (apart from a momentum--dependent field renormalization,
cf.\ the explicit expressions in the appendix). We consider it to be
essential, however, to use the modified form of the STI to fix the
gluon mass term, which is a truly relevant parameter of the effective
action.

Let us stress that the approximation we employ for the operators of type
b) does not imply that we are working 
with an effective action, which violates
the modified STI. In principle we could, for every value of $k \neq 0$,
determine the operators of type b) by solving the system of coupled
non--linear integral equations given by eq.\ (3.8), so that the resulting
effective action would be in accord with the modified STI. What we
actually do is to approximate the contributions of the operators of type b)
to the flow of the 2--point functions by the contributions of the
corresponding operators obtained from the standard STI, 
in the same way as we neglect the contributions of the operators
of type a).

It is important to note, that there exists a non--trivial quantitative
check for the self--consistency of the approximations employed. It is
based on the consistency of the ERGE with the $k$--derivative of the
modified STI in the absence of approximations, and will be discussed in more
detail in the next section. In particular, this self--consisteny condition
will be used to determine the range of scales $k$, where we consider
the approximations to be permissible.

The approximations employed here can be improved systematically by taking 
more and more operators of type a) into account, as well as by using
the operators of type b) corresponding to a (partial or complete)
solution of eq.\ (3.8). The purpose of the present paper is to see,
which results for the heavy quark potential will be obtained within
the ``minimal'' approach described above.

Let us now write down those terms in the $k$--dependent effective action,
which we take into account on the r.h.sides of the ERGEs for the 2--point 
functions. In a slight abuse of notation, we will denote this part of the
effective action by $\widehat{\Gamma}_k$. Firstly, we include the gluon and 
ghost 2--point functions themselves, which are given by the
momentum--dependent functions $f_{1,k}$ and $f_{2,k}$, respectively, and where 
we take a mass term for the gluons into account. Since we will consider
the Landau gauge later on, it is not necessary to introduce a
further momentum--dependent function in the longitudinal part of the
gluon 2--point function here. Secondly, we describe the
3--gluon, 4--gluon and ghost--gluon vertices of the type b) above
(determined from the standard STI), which
we take into account on the r.h.sides of the ERGEs for the 2--point
functions. Finally we include the external source terms containing
$K_{\mu}^a$, $L^a$ and $\bar{L}^a$, which are required to formulate 
the STI. 
These latter terms are constrained by two further identities satisfied by
$\widehat{\Gamma}_k$, which can be shown to be invariant under the RG
flow [11, 13, 14]: 
$$\partial_{\mu} {\delta \widehat{\Gamma}_k \over \delta K_{\mu}^a} =
{\delta
\widehat{\Gamma}_k \over \delta \bar{c}^a}\ \ \ , \eqno\mbox{(3.9a)}$$
$${\delta \widehat{\Gamma}_k \over \delta \bar{L}^a} = {1 \over \alpha}
\partial_{\mu} A_{\mu}^a \ \ \ . \eqno\mbox{(3.9b)}$$

\noindent Explicitly, then, our ansatz for $\widehat \Gamma_k$ reads
$$\widehat{\Gamma}_k(A, c, \bar{c}, K, L, \bar{L}) = $$
$${1 \over 2} \int_{p_1,p_2} \ A_{\mu}^a (p_1) \left \{ \left ( p_1^2
\ \delta_{\mu \nu} - p_{1 \mu} \ p_{1 \nu} \right ) f_{1,k} (p_1^2) +
{p_{1 \mu} \ p_{1 \nu} \over \alpha} + m_k^2 \ \delta_{\mu \nu} \right \}
A_{\nu}^a(p_2)$$
$${}+ i \ g_{k} \ f^a{}_{bc} \int_{p_1,p_2,p_3} \ f_{3, \mu \nu \rho}
(p_1, p_2, p_3) \ A_{\mu}^a(p_1) \ A_{\nu}^b(p_2) \ A_{\rho}^c(p_3)$$
$${}+ g_{k}^2 \ f^a{}_{bc} \ f^a{}_{de} \int_{p_1,\ldots,p_4} \ f_{4,\mu
\nu \rho \sigma}(p_1, p_2, p_3, p_4) \ A_{\mu}^b(p_1) \ A_{\nu}^c (p_2) \
A_{\rho}^d (p_3) \ A_{\sigma}^e(p_4)$$
$${}+ \int_{p_1,p_2} \ \bar{c}^a (p_1) \ p_1^2 \ f_{2,k}(p_1^2) \
c^a(p_2)$$
$${}+ i \ g_{k} \ f^a{}_{bc} \int_{p_1,p_2,p_3} \ \bar{c}^a(p_1) \ p_{1,\mu}
\ f_{2, k}(p_1^2) \ f_{2,k}^{-1}(p_2^2) \ A_{\mu}^b (p_2) \ c^c (p_3)$$
$${}+ i \int_{p_1,p_2} \ K_{\mu}^a(p_1) \ p_{1 \mu} \ f_{2,k}(p_1^2) \
c^a(p_2)$$
$${}- g_{k} \ f^a{}_{bc} \int_{p_1,p_2,p_3} \ K_{\mu}^a(p_1) \ f_{2,k}
(p_1^2) \ f_{2,k}^{-1}(p_2^2) \ A_{\mu}^b(p_2) \ c^c(p_3)$$
$${}- {g_{k} \over 2} \ f^a{}_{bc} \int_{p_1,p_2,p_3} \ L^a(p_1) \ c^b(p_2) 
\ c^c(p_3)$$
$${}- {i \over \alpha} \int_{p_1,p_2} \ \bar{L}^a (p_1) \ p_{1 \mu} \
A_{\mu}^a(p_2) \ \ \ , \eqno(3.10)$$
\noindent where
$$\int_{p_1,\ldots,p_n} \equiv \int \prod_{i=1}^n \left ( {d^4 p_i \over
(2 \pi )^4} \right
) \cdot (2 \pi)^4 \ \delta^4 \left ( \sum_{i=1}^n p_i \right ) \ \ \ .
\eqno(3.11)$$

\noindent The $k$--dependent 3--gluon and 4--gluon vertex
functions $f_{3, \mu \nu \rho}$ and $f_{4, \mu \nu \rho \sigma}$ are
given in terms of the gluon propagator function $f_{1,k}$ and the ghost
propagator function $f_{2,k}$ in the appendix. \par

In order to establish the connection with the heavy quark potential, we
introduce quark fields in the effective action. In the limit where the
quark mass tends to infinity, all quantum corrections corresponding to
diagrams which contain inner quark lines, are naively suppressed by
powers of the quark mass. On the other hand, the STI requires a 
momentum dependent quark--gluon vertex function, where the 
momentum dependence involves the function $f_{2,k}$ above. (Within the 
standard approach, the origin of this non-trivial contribution is due 
to UV divergent diagrams, which are not suppressed by the heavy quark 
mass.) After an eventual redefinition of the quark fields, such that their 
kinetic term is properly normalized,
the standard STI constrains the quark--gluon coupling in the effective
action to be of the form
$$ -\ g_k \ \lambda^a_{AB} \int_{p_1, p_2, p_3} \ \bar{\psi}^A (p_1)
\ \gamma_\mu \ f^{-1}_{2, k} (p_2^2) \ A^a_\mu (p_2) \ \psi^B (p_3) \ \ \ ,
\eqno(3.12) $$
where $\lambda$ denotes the SU(3) generators in the fundamental
representation. Here we have implicitely used the $Kc$--coupling from 
eq.\ (3.10) and
the fact that the quantum corrections to the vertices, which couple
external sources to the BRST--variations of the quark fields, are
suppressed by powers of the quark mass in the Landau gauge.

Let us now consider the 2--point function $V_k$ of the quark currents
$J^a_\mu$, given by
$$J_{\mu}^a = \ \lambda^a_{AB} \ \bar{\psi}^A \ \gamma_{\mu} \ \psi^B
\ \ \ . \eqno(3.13)$$
In the heavy quark limit, the only contribution to $V_k$ comes from the
dressed one--gluon exchange diagram, so from eq.\ (3.12) and the form of
the gluon propagator we obtain the current 2--point function
$$V_k(p^2) = {g_{k}^2 \over p^2 f_{1,k}(p^2) \ f_{2,k}^2(p^2)} \ \
\ . \eqno(3.14)$$

\noindent We have omitted the gluon mass term $m_k^2$ in our
definition of $V_k(p^2)$. In principle we could
have replaced the factor $p^2f_{1,k}(p^2)$ in the denominator of
$V_k(p^2)$, which is due to the gluon
propagator, by the factor involving $m_k^2$ and the infrared cutoff term.
However, once the integration of the ERGE is pursued down to $k = 0$,
the modified STI (3.8) turns into the standard  STI and
enforces $m_{k=0}^2 = 0$, and the infrared cutoff term vanishes
identically. Using eq.\ (3.14) to define $V_k(p^2)$, these effects
have been anticipated, and $V_k(p^2)$ corresponds more closely to
the physical function $V_0(p^2)$ already at finite $k$. \par

In the heavy quark limit $V_k(p^2)$ can be identified, up to trivial
factors, with the Fourier transform of the potential between quarks in 
quenched QCD. Although we will compute the gluon
and ghost propagator functions $f_{1,k}$ and $f_{2,k}$ individually by
integrating the
ERGEs, it is only the combination appearing in eq.\ (3.14), which has a
physical meaning. \par

A priori it may seem that the functions $f_{1,k}$ and $f_{2,k}$ can be 
changed at will by redefining the gluon, ghost and anti--ghost fields 
in the form of multiplication with momentum dependent functions. We have 
checked explicitely, however, that the STI restricts the relations between 
the vertices and the two--point functions such that finally all possible 
field redefinitions cancel in the expression for $V_k(p^2)$. 

As noted above and in refs.\ [11, 13, 14], $m_k^2$ in
$\widehat{\Gamma}_k$ is fixed in terms of the other
parameters in $\widehat{\Gamma}_k$ by the non--vanishing r.h.side of
eq.\ (3.8), where we use our ansatz (3.10) for $\widehat{\Gamma}_k$. 
After a consideration of terms
$\sim A \cdot c$ of eq.\ (3.8), one arrives at a non--linear relation
of the form
$$m_k^2 = {g_{k}^2 \over ( 4 \pi )^2} \, k^2 \ \mbox{ST}(k^2, m_k^2,
f_{1,k}, f_{2,k}) \ \ \ ,
\eqno(3.15)$$
\noindent where we have anticipated the Landau gauge and omitted the
gauge parameter
$\alpha$ on the r.h.side. The expression $\mbox{ST}(k^2, ... )$ 
involves momentum
integrations over a number of one--loop diagrams, in which the gluon and
ghost
propagators and hence the functions $f_{1,k}$ and $f_{2,k}$ appear, 
and will be given in the appendix.
Generally eq.\
(3.15) can be solved for $m_k^2$ only numerically, but in any case
$m_k^2$ is not an
independent parameter of $\widehat{\Gamma}_k$. \par

To determine the flow of the coupling parameter $g_k$, we will use the
ERGE for the $KcA$--vertex with vanishing momenta. Then it turns out,
that in the Landau gauge $g_k$ actually does not run at all
with $k$; it remains a
constant, which will be specified once and for all at the starting
point $k = \Lambda$. (It is straightforward, on the other hand, to
define a ``physical'' running coupling in terms of the dressed one--gluon
exchange diagram or the function $V_k$ of eq.\ (3.14), which runs with
$k$ due to the running of $f_{1,k}$ and $f_{2,k}$. Within our
approximation this running agrees to one--loop order with the usual 
perturbative running, as we have checked explicitely.) In the Landau gauge
the independent running parameters in $\widehat{\Gamma}_k$ are thus
only the two functions $f_{1,k}$ and $f_{2,k}$. The aim
will be the determination of these quantities by integrating the ERGEs
with suitable boundary conditions. \par

Finally we have to introduce and specify the IR cutoff term $\Delta
S_k$ in eq.\
(3.5). If we demand a reasonable behaviour of the full propagators
including the IR cutoff terms
which appear on the r.h.side of the
ERGE (3.7), we are lead to more sophisticated choices than an expression
of the form of
eq.\ (2.4) with $\Lambda \to \infty$. (Note that $R_{k, \mu \nu}$ and
$\tilde{R}_k$ have to vanish identically for $k \to 0$ in order that the
modified
STI (3.8) turns into the standard one (3.4). To this
end
$\Lambda$ has to be put equal to $\infty$ inside $R_{k, \mu \nu}$ and
$\tilde{R}_k$,
which is perfectly consistent as discussed in sect.\ 2.) First, we do not
want to
generate any poles on the real axis through the introduction of $\Delta
S_k$. This
enforces a dependence of $R_{k, \mu \nu}$ and $\tilde{R}_k$ on the
functions $f_{1,k}$
and $f_{2,k}$ such that they are proportional to $f_{1,k}$ and
$f_{2,k}$, respectively.
Now, however, it is no longer automatically guaranteed, that $\Delta
S_k$ serves as an
IR cutoff if, e.g., $f_{1, k}(p^2)$ vanishes for $p^2 \to 0$ (as we
might possibly
expect). This problem is overcome by writing the functions $f_{1,k}$ and
$f_{2,k}$ also
in the exponents. Taking the presence of the gluon mass term $m_k^2$
into account, our
choice of the cutoff functions thus reads
$$R_{k, \mu \nu}(p^2) = \left ( p^2 f_{1,k}(p^2) + m_k^2 \right )
{e^{-(p^2f_{1,k}(p^2) + m_k^2)/k^2} \over 1 -
e^{-(p^2f_{1,k}(p^2)+m_k^2)/k^2}}
\left ( \delta_{\mu \nu} - {p_{\mu} p_{\nu} \over p^2} \right )$$
$$+ \left ( {p^2
\over \alpha} + m_k^2 \right ) {e^{-(p^2/\alpha +
m_k^2)/k^2} \over 1 -
e^{-(p^2/\alpha + m_k^2)/k^2}} \, {p_{\mu}p_{\nu} \over p^2} \ \ \ ,
\eqno\mbox{(3.16a)}$$
$$\tilde{R}_k(p^2) = p^2f_{2,k}(p^2) \, {e^{-p^2f_{2,k}(p^2)/k^2} \over
1 - e^{-p^2
f_{2,k}(p^2)/k^2}} \ \ \ . \eqno\mbox{(3.16b)}$$

In order to obtain the full propagators, these terms have to be added to
the quadratic
terms $\sim A \cdot A$ or $\sim \bar{c} \cdot c$ in the action
$\widehat{\Gamma}_k$. Let us at this point remark a subtlety which has 
already been noted in ref.\ [14]: the modified
STI (3.8) determines, after its expansion to ${\cal
O}(A \cdot c)$, only
the longitudinal part of the gluon propagator and requires the presence
of a term of the
form $m_k^2 \ p_{\mu} p_{\nu}/p^2$. Then it is the condition of locality
of the effective
action, i.e.\ the need to cancel terms of the form $p_{\mu}p_{\nu}/p^2$,
which requires
the presence of the same mass term $m_k^2$ in the transverse part of the
gluon
propagator. Only after this consideration has been completed, we are
allowed to
approach the Landau gauge $\alpha = 0$. In ref.\ [14] we had checked
that $\alpha = 0$ is
preserved by the RG flow (and IR stable), and that the r.h.sides
of the ERGEs 
remain finite. Since this choice simplifies the calculations
considerably,
we will work in the Landau gauge throughout the rest of this paper. The
full gluon and
ghost propagators, as derived from $\Gamma_k = \widehat{\Gamma}_k +
\Delta S_k$, then
read
$${1 -
e^{-(p^2f_{1,k}(p^2) + m_k^2)/k^2} \over p^2 f_{1,k}(p^2) + m_k^2}
\left ( \delta_{\mu \nu} - {p_{\mu} p_{\nu} \over p^2} \right )
\eqno\mbox{(3.17a)}$$
\noindent and
$${1 - e^{-p^2f_{2,k}(p^2)/k^2} \over p^2f_{2,k}(p^2)} \ \ \ ,
\eqno\mbox{(3.17b)}$$
\noindent respectively. \par

\bigskip

\noindent {\bf 4. Procedure}

\medskip

Let us now turn to the computation of $f_{1,k}$ and $f_{2,k}$ by
integrating the ERGEs. The ERGE for $f_{1,k}$ can most
easily be
obtained by studying the terms quadratic in $A$ of the functional ERGE
 (3.7).
Its diagrammatic form is shown in fig.\ 1. Note that the 3-- and
4--gluon vertices
are more complicated than the ones of a classical Yang--Mills action
and are given in the appendix. The form of the 3--gluon
vertex coincides with the one used in the framework of SDEs [17] because
it corresponds to a special solution of the standard STIs 
[22] (up to a momentum--dependent field redefinition); we have
not been able to find the corresponding 4--gluon vertex elsewhere in
the literature.
The ERGE for $f_{2,k}$ is obtained from the terms $\sim
\bar{c}c$ of eq.\
(3.7), and is also shown in diagrammatic form in fig.~1.

The crossed circles in fig.\ 1 denote insertions of $\partial_k R_{k,
\mu \nu}$ resp.\
$\partial_k \tilde{R}_k$ according to the ERGE (3.7). Due to the
complicated
expressions (3.16) for these cutoff functions, i.e.\ their dependence on
$m_k^2$, $f_{1, k}$
and $f_{2,k}$, the $k$--derivatives $\partial_k$ acting on
$R_{k,\mu \nu}$ and $\tilde R_k$ have to be
decomposed into partial derivatives $\partial/\partial k$, \ $\partial
f_{1,k}/\partial k \cdot \partial/\partial f_{1,k}$, etc. \par

The system of ERGEs for $f_{1,k}$ and $f_{2,k}$ (in the Landau gauge), which 
is obtained from the diagrams of fig.\ 1, is thus of the form
$$\partial_k f_{1,k} = {g_{\Lambda}^2 \over (4 \pi)^2} \left (
h_1^0(k^2, m^2_k, f_{1,k},
f_{2,k}) + \partial_k f_{1,k} \ast h_1^1(k^2, m^2_k, f_{1,k},
f_{2,k}) \right .$$
$$\left . {}+ \partial_k f_{2,k} \ast h_1^2(k^2, m^2_k, f_{1,k}, f_{2,k}) +
\partial_k
m_k^2 \cdot h_1^3 (k^2, m^2_k, f_{1,k}, f_{2,k}) \right ) \ \ \ ,
\eqno\mbox{(4.1a)}$$

$$\partial_k f_{2,k} = {g_{\Lambda}^2 \over (4 \pi)^2} \left (
h_2^0(k^2, m^2_k, f_{1,k},
f_{2,k}) + \partial_k f_{1,k} \ast h_2^1(k^2, m^2_k, f_{1,k},
f_{2,k}) \right .$$
$$\left . {}+ \partial_k f_{2,k} \ast h_2^2(k^2, m^2_k, f_{1,k}, f_{2,k}) +
\partial_k m_k^2 \cdot h_2^3 (k^2, m^2_k, f_{1,k}, f_{2,k}) 
\right ) \ \ \ . \eqno\mbox{(4.1b)}$$

The evaluation of the expressions $h_i^j$ requires a numerical
computation of the one--loop integrals appearing in fig.~1. The $\ast$
denotes corresponding convolutions with respect to the loop momentum. It
is evident from
eqs.\ (4.1), that we still need the knowledge of $\partial_k m_k^2$ in
order to
determine $\partial_k f_{1,k}$ and $\partial_k f_{2,k}$. $\partial_k
m_k^2$
can be obtained from the same diagrams in fig.\ 1, which are relevant for
$\partial_k
f_{1,k}$, in the limit of vanishing external momentum. Again this gives
us an
expression of the form
$$\partial_k m_k^2 = {g_{\Lambda}^2 \over (4 \pi )^2} \left (
h_3^0(k^2, m^2_k, f_{1,k},
f_{2,k}) + \partial_k f_{1,k} \ast h_3^1 (k^2, m^2_k, f_{1,k}, f_{2,k})
\right .$$
$$\left . {}+ \partial_k f_{2,k} \ast h_3^2 (k^2, m^2_k, f_{1,k},
f_{2,k}) +
\partial_k m_k^2 \cdot h_3^3 (k^2, m^2_k, f_{1,k}, f_{2,k}) \right ) \ \ \ .
\eqno(4.2)$$

Alternatively, we can obtain a similar expression by differentiating
eq.\ (3.15) with respect to $k$. We thus have two different equations at
our disposal to complete the system (4.1). In the absence of
approximations, the two different completed systems 
would yield exactly the same results, as follows from the compatibility of
the full equations (3.7) and (3.8). 
Since the r.h.sides of the eqs.\ (4.2) and (3.15) involve our
ansatz (3.10) for $\widehat{\Gamma}_k$ rather than the full effective
action, however, we obtain different results. (Nevertheless, they agree on 
the one--loop level [14].) The use of $\widehat{\Gamma}_k$ instead
of the full effective action involves two different kinds of approximations
related to the operators of type a) and b), respectively (see the previous
section). 

The compatibility of the two completed systems of equations
considered above turns out to be sensitive to both kinds of approximations.
As in ref.\ [14],
we will use the difference between the corresponding results as a quantitative
measure of the inadequacy of the approximations (for the respective scale
$k$). Here we concentrate on the effect of the approximations on the
gluon and ghost propagator functions: We will take a certain discrepancy
(in fact 10\%) in the results for $\partial_k f_{1,k}$ or 
$\partial_k f_{2,k}$ at some value of the momentum as an indication that
the approximations have a considerable influence on the contributions to
the flow of  $f_{1,k}$ or $f_{2,k}$.

We are of course aware of the fact that the described consistency check
merely yields a necessary condition for the validity of the approximation,
not a sufficient one. There exists, however, another a--posteriori
justification for our procedure: The results we obtain will show a
remarkable degree of universality in the sense of ref.\ [23] (see the
next section) as long as $k > k_0$, where the scale $k_0$ is determined
from the above consistency condition. Since universality is only expected
to hold when gauge symmetry is taken into account properly, and no
significantly contributing operators are neglected, this property
provides a strong indication of the consistency of our procedure.

In practice we will use the form of eq.\ (4.2) as derived from
the ERGE (3.7) in order to complete the system of linear equations for 
$\partial_k f_{1,k}$ and $\partial_k f_{2,k}$. The other system of equations,
with eq.\ (4.2) taken from the derivative of eq.\ (3.15), will be employed
for the consistency check at every scale $k$. \par

Given the ERGEs for $f_{1,k}$ and $f_{2,k}$, we still need suitable boundary 
conditions in order to proceed with the
numerical integration. These boundary conditions can be chosen such
that $\widehat{\Gamma}_\Lambda$ resembles as close as possible the classical 
(bare) Yang--Mills action (3.1). One would thus choose, at
$k = \Lambda$, $f_{1, \Lambda}(p^2) = f_{2,
\Lambda}(p^2) = 1$ and $g_{\Lambda}$ not too large in order to start in
the perturbative regime. It is possible, however, to improve the form of
the effective action at the starting point $k = \Lambda$. By definition,
$\widehat{\Gamma}_\Lambda$ should include all quantum effects involving
internal momenta $p^2 \geq \Lambda^2$. On the one hand these are
certainly small for a small coupling $g_{\Lambda}$, on the other hand it
is fairly straightforward to include them to one--loop order. To this end
one has to calculate the one--loop diagrams, which contribute to the
gluon and ghost propagator, with IR cutoff functions $R_k$ as in eqs.\
(3.16) and $k = \Lambda$. (Inside these cutoff functions we can set,
to one--loop order, $f_{1,k}(p^2) = f_{2,k}(p^2) = 1$ and $m_k^2 = 0$). The
diagrams have to be regularized in the UV (e.g.\ dimensionally) and
renormalized such that, e.g., the renormalization conditions

$$f_{1, \Lambda}(0) = 1 \quad , \quad f_{2, \Lambda}(0) = 1 \eqno(4.3)$$

\noindent are satisfied. As a result one obtains
$$
f_{1, \Lambda}(p^2) = 1 + \delta f_1^{\mbox{\scriptsize 1--loop}}(p^2)
\ \ \ , $$
$$
f_{2, \Lambda}(p^2) = 1 + \delta f_2^{\mbox{\scriptsize 1--loop}} (p^2)
\eqno(4.4)
$$
\noindent where the one--loop contributions $\delta 
f^{\mbox{\scriptsize 1--loop}}_i$ are given in the
appendix. For large $p^2$ the functions $\delta 
f^{\mbox{\scriptsize 1--loop}}_i$ behave asymptotically as
$$
\delta f_1^{\mbox{\scriptsize 1--loop}}(p^2)  \sim {g_{\Lambda}^2 \over (4
\pi)^2} \cdot {13 \over 2} \ln \left
({p^2 \over \Lambda^2} \right ) \ \ \ , $$
$$
\delta f_2^{\mbox{\scriptsize 1--loop}}(p^2) \sim {g_{\Lambda}^2 \over (4
\pi)^2} \cdot {9 \over
4} \ln \left ( {p^2 \over \Lambda^2} \right ) \ \ \ .
\eqno(4.5)
$$

\noindent What we have achieved now is an appropriate behaviour of these
functions even for momenta $p^2 \gg \Lambda^2$, far above the starting
scale, provided  $g_{\Lambda}^2/(4 \pi)^2 \cdot \ln (p^2/
\Lambda^2) \ll 1$. Now $V_k(p^2)$ of eq.\ (3.14) has the
appropriate logarithmic $p^2$--dependence even for $p^2 \gg \Lambda^2$,
corresponding to a one--loop RG improvement of the ``physical'' coupling
constant:
$$ V_{k=\Lambda}(p^2 \gg \Lambda^2) \sim
{\displaystyle g_{\Lambda}^2 \over \displaystyle {p^2
\left (1+{11 g_{\Lambda}^2 \over (4\pi)^2} \ln \left ({p^2 \over
\Lambda^2} \right ) \right )} } \ \ \ .  \eqno(4.6)$$

As boundary conditions for the integration of the ERGEs we thus use
propagator functions
$f_{i, \Lambda}$ according to eqs.\ (4.4), the numerical
solution of
eq.~(3.15) for $m_\Lambda^2$, and values for $g_{\Lambda}$ (as the only free
parameter) from $g_{\Lambda} = 2.0$ down to $g_{\Lambda} = 1.4$. \par

It is clear that, in order to integrate the ERGEs (4.1) for the
functions
$f_{1,k}$ and $f_{2,k}$, numerical methods have to be
employed. To this end we
need a parametrization of these functions, which allows for a reliable
fit for all values
of the scale $k$. Since these functions always behave like $f_{i,k}(p^2
\to \infty ) \to 1 + \delta f_i^{\mbox{\scriptsize 1--loop}}$, the
following parametrization turns out to be convenient:
$$f_{i,k}(p^2) = 1 - \sum_j {\alpha_j^i \over 1 + \gamma_j^i \, p^2}
+ \delta f_i^{\mbox{\scriptsize 1--loop}}({p^2})
\eqno(4.7)$$
\noindent with up to 6 pairs of parameters $\alpha_j^i$, $\gamma_j^i$
for $i = 1,2$,
respectively. \par

Next we have to diagonalize eqs.\ (4.1) in the space of momenta $p^2$.
To this end we
evaluate the functions $h_i^j$ in eqs. (4.1) and (4.2) for a number of
momenta $p^2 =
p_j^2$ with up to 35 different values for $p_j^2$ in the range
$10^{-1} k_0^2$ to
$10 \Lambda^2$, where $k_0^2$ is determined dynamically (see below).
(It turns out to be convenient to space the momenta $p_j^2$ equally on a
logarithmic scale.)
By solving the system of linear algebraic equations we then
obtain
expressions for $\partial_k f_{i,k}(p_j^2)$ and $\partial_k m_k^2$.
One step in the
integration of the ERGE then amounts to a computation of
$$f_{i,k+\Delta k}(p_j^2) = f_{i,k}(p_j^2) + {\Delta k}
\cdot \partial_k
f_{i,k}(p_j^2) \eqno(4.8)$$
\noindent with $|\Delta k/k| < 3 \cdot 10^{-2}$, and a subsequent
determination of the new
fit parameters $\alpha_j^i$, $\gamma_j^i$ of eq.\ (4.7). \par

The new value of the parameter $m_k^2$ gets determined by a new
numerical solution of
eq.~(3.15) rather than by the numerical integration of $\partial_k
m_k^2$; since it
is a relevant parameter (the only one), for $k^2 \ll
\Lambda^2$ accumulated theoretical and numerical errors would easily
generate a $m_k^2$ of ${\cal O}(\Lambda^2)$ instead of ${\cal O}(k^2)$,
as it should be according to eq.\ (3.15). \par

Of course we have varied many details of the numerical procedure in
order to test the
robustness of the final results. An upper limit of $\sim$ 35
different values for
$p_j^2$ is dictated by the finite amount of available computing time,
and this also keeps us from
choosing arbitrarily small values for the gauge coupling $g_{\Lambda}$
at the starting
point $k^2 = \Lambda^2$, since then the ERGEs would have to be
integrated over
arbitrarily many orders of magnitude of the scale $k^2$ in order to arrive
at the
physically interesting non--perturbative regime.
\par \bigskip

\noindent {\bf 5. Results}

\medskip

Our results can most easily be represented in terms of a function $F_k(p^2)$,

$$F_k(p^2) = {1 \over g_{\Lambda}^2} \ f_{1, k}(p^2) \ f_{2,k}^2(p^2) \
\ \ , \eqno(5.1)$$

\noindent which is simply related to the potential $V_k(p^2)$ in eq.\ (3.14):

$$ V_k(p^2) = \frac{1}{p^2 F_k(p^2)} \ \ \ . \eqno(5.2) $$

\noindent In figs.\ 2--4 we
show our results for $F_k(p^2)$ for a bare coupling $g_{\Lambda} = 1.4$.
We plotted this function for different values of $k^2$, $k^2 = \Lambda^2$,
$10^{-1} \Lambda^2$, $10^{-2} \Lambda^2$, $10^{-3} \Lambda^2$,
$k^2 = {k_0}^2 = 7.43 \cdot 10^{-5} \Lambda^2$, as full lines, and 
for some still smaller value 
$k^2 = \hat k^2 = 3.68 \cdot 10^{-5} \Lambda^2$ as a dashed line. $p^2$ is
given in units of $\Lambda^2$; the maximal value of $p^2$ is
$\Lambda^2$, $10^{-1} \Lambda^2$ and, in order to resolve the small
$p^2$ region, $10^{-2} \Lambda^2$ in figs.~2 to 4, respectively. The
curves with decreasing values at $p^2 = 0$ correspond to decreasing
values of $k^2$. 
We see that, for small enough $k^2$ and small enough $p^2$, $F_k(p^2)$
approaches a form $\sim p^2$ near the origin, which corresponds to
a $1/p^4$--behaviour of the potential $V_k(p^2)$. 

Just before $F_k(p^2)$ becomes $0$ for $p^2 \to 0$, however, the two
different methods of evaluating $\partial_k m_k^2$ (cf.\ eq.\ (4.2) and
the discussion below) lead to different results for $\partial_k f_{1,k}$
or $\partial_k f_{2,k}$, with a relative difference of $\sim 0.1$. We 
denoted the corresponding scale by ${k_0}$. For slightly smaller values 
of $k$ the equation for the gluon mass term $m_k^2$ (3.15) possesses no 
longer a solution for $m_k^2$. 
\par

The incompatibility of the RG flow with the modified STIs 
(3.8) below $k_0$ indicates, that our approximation becomes
unreliable in this regime: Numerically different results for 
$\partial_k f_{1,k}$ or $\partial_k f_{2,k}$ show, that the neglected 
contributions play an important role on the r.h.side of the ERGE for the 
gluon or ghost propagator, respectively, and that
our ansatz is too restrictive in this regime. \par

The gluon mass term $m_k^2$ itself, which was determined by eq.\ (3.15) 
to be of ${\cal
O}(\Lambda^2)$ at the starting point $k = \Lambda$, does not show an
unusual behaviour for
$k \sim {k_0}$. $m_k^2$ is always of ${\cal O} (k^2)$ or, more precisely,
$m_k^2 \sim -0.6 \cdot k^2$ for $k \to {k_0}$. \par

Thus we cannot have confidence in the form of the RG flow for $k <
{k_0}$. However, generally the RG flow modifies $F_k(p^2)$ only for $p^2
\;\parbox{0.35cm}{$\displaystyle\stackrel{\textstyle <}
{\sim}$}\; k^2$ (cf. figs.~2--4), hence our result for $F_{k_0}$
 is nevertheless trustworthy for $p^2 \;\parbox{0.35cm}{$\displaystyle
\stackrel{\textstyle >}{\sim}$}\; {k_0}^2$. Our results also 
indicate that, within a more general parametrization of $\hat
\Gamma_k$, $F_k(0)$ approaches $0$ for $k^2 \to 0$: In figs.\ 2--4 we have 
shown the function $F_k(p^2)$ for some value $k^2 = \hat k^2 < {k_0}^2$ 
as a dashed line, and we see that the decrease of $F_k(p^2)$ at $p^2 = 0$ 
continues. For still smaller values of $k^2$, however, the integration 
of the system of differential equations becomes numerically unstable 
and does no longer allow to obtain reliable results. \par

In any case, already our results for $F_{{k_0}}(p^2)$ for $p^2 \;
\parbox{0.35cm}{$\displaystyle\stackrel{\textstyle >}{\sim}$}\; 
{k_0}^2$ strongly indicate a form $\sim p^2$ near the origin, and the
trustworthy range of $p^2$ may be sufficient for some phenomenological
investigations involving, e.g., heavy quarks. \par

It might be interesting to get some feeling for the orders of
magnitude of the different scales involved. Up to now all scales are
only defined 
relative to the starting scale  $\Lambda$. In order to relate this scale
to a physical scale one can try to compare our result for $V_{k_0}(p^2)$,
 for $p^2 \;\parbox{0.35cm}{$\displaystyle\stackrel{\textstyle >}
{\sim}$}\; {k_0}^2$, with a phenomenological parametrization of this
potential, which contains dimensionful parameters known in units of
MeV. A convenient form of such a parametrization is given by
Richardson [24], which reads in our convention
$$V_R(p^2) = {48 \pi^2 \over {(33-2N_f)p^2 
\ln\left (1+p^2/\Lambda_R^2 \right )}}
\eqno(5.3)$$
\noindent with $\Lambda_R \cong 400$ MeV, and for a rough comparison
with our results (in pure SU(3) Yang--Mills theory) we may set $N_f = 0$. In
terms of our function $F_k(p^2)$ of eq.~(5.1) this ansatz reads
$$F_R(p^2) = {11 \ln \left (1+p^2/\Lambda^2_R \right ) 
\over 16 \pi^2 } \ \ \ . \eqno(5.4)$$
\noindent We performed a fit of eq.\ (5.4) to our function $F_{k_0}(p^2)$ for
the region of momenta $k_0^2 \ll p^2 \ll \Lambda^2$ to determine the ratio
$\Lambda / \Lambda_R$. The result is shown in
figs.\ 5 and 6, where we plotted $F_{k_0}(p^2)$ as a full  line and 
$F_R(p^2)$, eq.~(5.4), as a dashed line, against $p^2$ in GeV$^2$. The
optimized relation between $\Lambda$ and $\Lambda_R$ is such that
$\Lambda = 12.1$ GeV, which implies that the scale ${k_0}$, where our
approximation ceases to be appropriate, is $\sim 104$ MeV. 
One can see that our potential is surprisingly close
to the phenomenological one of Richardson, both in the perturbative
(fig.\ 5) and in the nonperturbative (fig.\ 6) regime. Significant deviations 
show up only for momenta $p^2 \;\parbox{0.35cm}{$\displaystyle
\stackrel{\textstyle <}{\sim}$}\; {k_0}^2$. \par

Let us now look at the dependence of our results on the bare coupling
$g_{\Lambda}$. We find that also for larger values of $g_{\Lambda}$, $F_k(0)$ 
becomes small for small $k$, until our consistency condition indicates the 
inadequacy of the approximation at some scale ${k_0}$, with
${k_0}^2/\Lambda^2 = 6.13 \cdot 10^{-4}$, $2.69 \cdot 10^{-3}$,
and $7.98 \cdot 10^{-3}$ for $g_{\Lambda} =  1.6$, $1.8$ and $2.0$,
respectively. In fig.\ 7 we compare the different
results for $F_{k_0}(p^2)$ with $g_{\Lambda} = 1.4$ (full line),
$g_{\Lambda} = 1.6$ (long--dashed line), $g_{\Lambda} = 1.8$
(short--dashed line) and $g_{\Lambda} = 2.0$ (dotted line).
In each case we determined the physical value of the starting scale by
a fit to eq.\ (5.4), and the results for  $\Lambda$ are given by 12.1 GeV,
 4.8 GeV, 2.6 GeV and 1.7 GeV for $g_{\Lambda} = 1.4,  1.6, 1.8$ and $2.0$, 
respectively. From the values of $k_0^2/\Lambda^2$ given above, we then
obtain the corresponding values for $k_0$, ${k_0} = 104$ MeV, 119 MeV, 
134 MeV and 151 MeV. We see that the curves in fig.\ 7
coincide remarkably well, which assures us that the numerical 
method we use for the integration of the ERGEs does not accumulate numerical 
errors. The approximate independence of the curves on $g_{\Lambda}$ or
$\Lambda$ even for momenta larger than the respective starting scale
is due to the one--loop improvement of the starting action $\widehat
\Gamma_{\Lambda}$ (or of the functions $f_{1,\Lambda}$ and $f_{2,\Lambda}$
according to eq.\ (4.4)). In fig.\ 8  we show the same curves  for 
$F_{k_0}(p^2)$ for the  different  values  of $g_{\Lambda}$ or $\Lambda$, 
which have  been obtained without  the  one--loop improvement. \par

Figs.\ 7 and 8 lead to a series of important observations: First, as we
see most clearly from fig.\ 8, the curves are essentially unchanged
compared to their form at the respective starting scale $\Lambda$ for
momenta $p^2 \gg \Lambda^2$, as they should be. In particular, the curve
for $\Lambda = 1.7$ GeV ($g_\Lambda = 2.0$) in fig.\ 7 is given in the
perturbative region essentially by the one--loop form (4.4), so
comparison with, e.g., the curves for $\Lambda = 12.1$ GeV ($g_\Lambda
= 1.4$) in both figures shows, that the integration of the ERGEs
correctly reproduces the one--loop form in the perturbative domain.
More importantly, we find universal behaviour in the non--perturbative
domain: For momenta $p^2 \ll \Lambda^2$, i.e. below the respective
starting scales, all curves merge in one, as is most impressively
demonstrated by fig.\ 8, and the form of the resulting curve does not
depend on the boundary condition we use.
This independence of physics at $p^2 \ll \Lambda^2$ with respect to
variations of the starting scale $\Lambda$ and to the inclusion of
irrelevant operators at $\Lambda$ (as is the case for the one--loop
improved boundary conditions) is what we call universality here, in
accordance with ref.\ [23].

It is easy to see that the relation between  $g_{\Lambda}$ and the
starting scale $\Lambda$, which has been obtained empirically above,
agrees with the one--loop running of $g_{\Lambda}$. On the other hand the
running of $g_{\Lambda}$ with $\Lambda$ is not yet correct to
two--loop accuracy. However, with our ansatz for $\widehat{\Gamma}_k$
only a  subset  of all possible two--loop diagrams is included within a
perturbative  expansion of the ERGEs, hence an agreement at this level
would have been mere luck. \par

One may ask whether the behaviour of $F_{k_0}(p^2)$ for $p^2 \to 0$
is due to the form of $f_{1,k_0}(p^2)$ or both
 $f_{1, k_0}(p^2)$ and $f_{2,k_0}(p^2)$. To this end we plot in
fig.~9 both functions $f_{1,k_0}(p^2)$ and $f_{2,k_0}(p^2)$ (with $p^2$
rescaled as
before) for $g_{\Lambda} = 1.4$. We see that both functions become
 small for $p^2 \to 0$, but it would be grossly
misleading (and unphysical) to identify the gluon propagator
function $(p^2
f_1(p^2))^{-1}$ with the heavy quark potential and to neglect the dependence
of $V(p^2)$ of eq.~(3.14) on $f_2(p^2)$, as it is done if the ghost part of
the action
is neglected, and the STIs are not properly taken into
account. \par
\bigskip

\noindent {\bf 6. Discussion and Outlook}

\medskip
In this paper we have performed the integration of the ERGEs for
non--abelian gauge theories, with the bare action as the only input. 
We concentrated on pure SU(3) Yang--Mills theory (or quenched QCD),
and the momentum dependence of the gluon and ghost propagators; these
quantities determine the heavy quark potential. \par

The neglect of certain contributions on the r.h.sides of the ERGEs for
the 2--point functions turned the system of ERGEs into a closed system
of differential equations, which we integrated numerically. A priori the
validity of the approximation is difficult to judge in the strong
coupling regime, where it amounts to the neglect of contributions of the
same order in the coupling constant as the ones maintained. Here the use 
of modified STIs plays an important role: Due to their
validity parts of the $k$--dependent effective action can be determined
in two different ways, namely either directly through these identities or
by integrating the ERGEs. Without approximations both methods give the
same results, but in the presence of approximations a possible difference 
between the corresponding results can be used to estimate the error 
induced by the approximations. In our case it was the gluon mass term, which 
was obtained in two different ways. We used a 10\%--deviation in 
$\partial_k f_{1,k}$ or $\partial_k f_{2,k}$, induced by the 
different methods to obtain $\partial_k m_k^2$, as an indication for the
breakdown of our approximation near the corresponding scale $k_0$. 
Accordingly, our result for $V(p^2) \equiv V_{k_0}(p^2)$ is only 
consistent for $p^2 \;\parbox{0.35cm}{$\displaystyle\stackrel
{\textstyle >}{\sim}$}\;{k_0}^2$, where we estimated $k_0 \sim 100$ MeV. 
Our main findings are the
strong indication of a $1/p^4$--behaviour of $V(p^2)$ in the trustworthy
regime of $p^2$, and the fact that the full form of $V(p^2)$ happens to
be close to the one proposed by Richardson [24]. \par

In the second section of this paper we discussed the formal relation
between ERGEs and
SDEs, with the result that effective actions, which solve SDEs in the
presence of an IR
cutoff $k$, can be considered as quasi--fixed points of the ERGEs.
Although this formal
relation is no longer exact in the presence of truncations of
effective actions, it
is thus not astonishing, that the behaviour of the gluon propagator,
which has
been derived here using the ERGEs, has also been shown to be a solution
to SDEs in
the Landau gauge [17]. We believe, however, that the present method
possesses a number
of advantages compared to the formalism of the SDEs: Firstly, the
regularization of both
ultraviolet and infrared singularities in the context of the SDEs
generates notorious
technical difficulties, which require, to some extent, ad hoc
prescriptions, the
influence of which on the final result is difficult to control. The
present method is,
in contrast, free of problems related to UV or IR singularities by
construction. \par

Secondly, a priori both methods require a suitable truncation of the
effective action. Within
the present method, however, it is technically simpler to include more
and more terms; for the present investigation, e.g., we already included 
the contributions
of the 4--gluon vertex and the ghost fields, which were neglected in
the SDE approach [17]. We have stressed that the detailed form of the
dressed one--gluon exchange diagram between heavy quarks, which involves
the ghost propagator function at the quark--gluon vertex due to the
STIs, differs considerably from the form of the
inverse gluon propagator alone (in the Landau gauge), which is not a
physical quantity by itself. Moreover, it is technically feasible in our
formalism to include further terms in the effective action without too much 
effort. \par

Finally, within the SDE approach one is only able to check the
self--consistency of a
particular ansatz; it is practically impossible to prove rigorously,
that the chosen
ansatz is the only possible one. The integration of the ERGEs,
on the other
hand, always gives a unique answer, namely the one which is
smoothly
connected to the corresponding bare action, and to perturbation theory
in the UV in the
case of asymptotic freedom. Actually, our discussion in section 2 sheds
some light on the cases where different solutions of the SDEs exist:
on the one hand solutions of the SDEs can be considered as fixed points
of the ERGEs for $k \to 0$, on the other hand it is generally
not clear,
whether one has obtained an IR stable or an UV stable fixed point. Since
only the IR
stable solutions can also be obtained by the integration of the ERGEs
 towards
$k \to 0$, one is lead to the conclusion, that only these are actually
physically
relevant, which is certainly difficult to check within the SDE approach
alone. \par

Many possibilities exist in order to extend the present approach in the
future: firstly, still more terms can be included in the effective action, 
and collective fields, e.g.\ for $F^a_{\mu \nu} F^a_{\mu \nu}$, can be  introduced along the lines of refs.\ [7, 8] in
order to reach renormalization scales $k$ below ${k_0}$,
and to get a still better understanding of pure Yang--Mills theory
free of systematic
uncertainties for $p^2 < {k_0}^2$. Secondly, the quark sector of QCD can
be introduced as well. First efforts
in this direction have already been made in ref.\ [8]; there, however,
the effects of
the gluons have been guessed and parametrized in a phenomenological
manner. Now the
integration of the ERGEs for the full quark gluon system is
within reach
(possibly along the lines proposed in ref.\ [25]), which will allow to
study many
phenomenologically interesting systems like heavy $q\bar{q}$--bound
states, $q\bar{q}$--condensates
and the meson sector on the basis of just the bare QCD
Lagrangian.   \par

\vspace{0.9cm}

\noindent{\Large \bf Appendix}

\vspace{0.3cm}

In this appendix we list some of the functions which are needed in the
actual computations. We begin with the 3-- and 4--gluon vertices
employed in our ansatz (3.10) for $\widehat{\Gamma}_k$. They are chosen
in such a way as to fulfill the standard STIs in
a ``minimal'' fashion. Explicitly they read
$$ f_{3, \mu \nu \rho} (p, q, r) \;\: = \;\: -\, \frac{1}{6 f_{2,k}
(p^2) f_{2,k}(q^2) f_{2,k}(r^2)} \: \times $$
$$ \Bigg\{ \left( q_{\mu}f_{g,k}(q^2) - r_{\mu}f_{g,k}(r^2) \right)
\delta_{\nu\rho}
\: + \: \left( r_{\nu}f_{g,k}(r^2) - p_{\nu}f_{g,k}(p^2) \right)
\delta_{\rho\mu} $$
$$ {}+ \left( p_{\rho}f_{g,k}(p^2) - q_{\rho}f_{g,k}(q^2) \right)
\delta_{\mu\nu} $$
$$ {}+ \left(q_{\mu} - r_{\mu}\right)\frac{f_{g,k}(q^2) - f_{g,k}(r^2)}
{q^2 - r^2}\left(r_{\nu}q_{\rho} - q\cdot r \, \delta_{\nu\rho} \right)$$
$$ {}+ \left(r_{\nu} - p_{\nu}\right)\frac{f_{g,k}(r^2) - f_{g,k}(p^2)}
{r^2 - p^2}\left(p_{\rho}r_{\mu} - r\cdot p \, \delta_{\rho\mu} \right) $$
$$ \left. {}+
\left(p_{\rho} - q_{\rho}\right)\frac{f_{g,k}(p^2) - f_{g,k}(q^2)}
{p^2 - q^2}\left(q_{\mu}p_{\nu} - p\cdot q \, \delta_{\mu\nu} \right)
\right\} \eqno(\mbox{A}.1) $$
\noindent and
$$ f_{4, \mu \nu \rho \sigma} (p,q,r,s) \;\: = \;\: \frac{1}{4 f_{2,k}
(p^2) f_{2,k}(q^2) f_{2,k}(r^2) f_{2,k}(s^2)} \: \times $$
$$ \left\{ f_{g,k}((p+q)^2)\delta_{\mu\rho}\, \delta_{\nu\sigma}
\: - \: 2 \, \frac{f_{g,k}(p^2) - f_{g,k}(s^2)}{p^2 - s^2}
\left( p_{\sigma}s_{\mu} - p\cdot s \, \delta_{\sigma\mu} \right)
\delta_{\nu\rho} \right. $$
$$ {}- 2 \, \frac{f_{g,k}((p+q)^2) - f_{g,k}(p^2)}{(p+q)^2 - p^2}
\left( 2p_\nu + q_\nu \right)
\left(p_{\rho}\delta_{\sigma\mu} - p_{\sigma}\delta_{\mu\rho}\right) $$
$$ {}+ \frac{2}{(r + s)^2 - s^2}\left( \frac{f_{g,k}(p^2) - f_{g,k}
((r+s)^2)}{p^2 - (r+s)^2}
\: - \: \frac{f_{g,k}(p^2) - f_{g,k}(s^2)}{p^2 - s^2} \right) $$
$$ {}\times \left( 2p_\nu + q_\nu \right) \left( 2s_\rho +r_\rho \right)
\left( p_{\sigma}s_{\mu} - p\cdot s \,
\delta_{\sigma\mu} \right) \Bigg\} \ \ \ , \eqno(\mbox{A}.2) $$
\noindent where
$$ f_{g,k} (p^2) = f_{1,k} (p^2) f_{2,k}^2 (p^2) \ \ \ . \eqno(\mbox{A}.3)
$$
Observe that we have not symmetrized the expression (A.2) for the
four--gluon vertex with respect to
momenta and indices, in order to keep it to a reasonable length.

Next we turn to the modified STI for the gluon
mass term, eq.\ (3.15). In diagrammatic form it is given by
a number of one--loop diagrams. The integrations over the angular variables
of the loop momenta can be performed, and one is left with
integrals over $t \equiv q^2$. After a careful consideration of the
limit $\alpha \to 0$, corresponding to the Landau gauge, the following 
expression for the function ST results:
$$ \mbox{ST}(k^2, m_k^2, f_{1,k}, f_{2,k}) \;\: = \;\: $$
$$ \frac{3}{4 f^2_{2,k}(0)} \, \int\limits_0^\infty dt \, \frac{t}{k^2} 
\left\{ -e^{-t f_{2,k}(t)/k^2} \left[ 3 f_{1,k}(t) \, \frac{1 - e^{-(t
f_{1,k}(t) + m_k^2)/k^2}}{t f_{1,k}(t) + m_k^2} \right. \right. $$
$$ \left. {}+ \left( 3 - 4 \frac{t}{k^2}
\frac{\partial}{\partial t} \Big(t f_{2,k}(t) \Big) \right) f_{2,k}(t) \, 
\frac{1 - e^{-t f_{2,k}(t)/k^2}}{t f_{2,k}(t)} \right] $$
$$ {}+ 3 e^{-(t f_{1,k}(t) + m_k^2)/k^2} \left[ \left( 4 - 4 \frac{t}{k^2}
\frac{\partial}{\partial t} \Big(t f_{1,k}(t) \Big) \right) f_{1,k}(t) \, 
\frac{1 - e^{-(t f_{1,k}(t) + m_k^2)/k^2}}{t f_{1,k}(t) + m_k^2} \right. $$
$$ \left. {}+
f_{2,k}(t) \, \frac{1 - e^{-t f_{2,k}(t)/k^2}}{t f_{2,k}(t)} \right] 
\: - \: 6 \, \frac{e^{-(t f_{1,k}(t) + m_k^2)/k^2}}{f_{2,k}^2(t)} \, 
t \frac{\partial}{\partial t} \Big(f_{1,k}(t) f_{2,k}^2(t) \Big) \: \times $$
$$ \left[
\left( 1 + \frac{t}{k^2} \frac{\partial}{\partial t} \Big(t f_{1,k}(t) \Big)
- \frac{2}{f_{2,k}(t)} \frac{\partial}{\partial t} \Big(t f_{2,k}(t) \Big) 
\right)
\frac{1 - e^{-(t f_{1,k}(t) + m_k^2)/k^2}}{t f_{1,k}(t) + m_k^2} \right. $$
$$ {}- t \frac{\partial}{\partial t} \Big(t f_{1,k}(t) \Big) \,
\frac{1 - \Big( 1 + (t f_{1,k}(t) + m_k^2)/k^2 \Big) e^{-(t f_{1,k}(t) 
+ m_k^2)/k^2}}{\Big( t f_{1,k}(t) + m_k^2 \Big)^2} \Bigg]
\Bigg\} \ \ \ . \eqno(\mbox{A}.4) $$

Finally we give the one--loop expressions $\delta 
f^{\mbox{\scriptsize 1--loop}}_i$ for the gluon and ghost 
2--point functions as needed in (4.4) for the improved boundary conditions.
They are calculated with the IR cutoff functions $R_k$ from eqs.\ (3.16)
(with $f_{1,k}$ and $f_{2,k}$ set equal to one and $m_k^2 = 0$) and obey the
renormalization conditions (4.3). The results are, again in the Landau
gauge,
$$ \delta f_1^{\mbox{\scriptsize 1--loop}}(p^2) \;\: = \;\: 
\frac{3 g_\Lambda^2}{(4\pi)^2} \left\{ \frac{13}{6} \Big(\ln (
x/2) + C \Big) \: - \: \frac{131}{144}
\: + \: \frac{27 x^2 + 88 x - 20}{24 x^3} \right. $$
$$ {}- \frac{x^4 - 8 x^3 + 9 x^2 - 20 x - 6}{12 x^3} \, e^{-x} \:
+ \: \frac{x^4 - 16 x^3 - 16 x^2 - 56 x + 4}{12 x^3} \, e^{-x/2} $$
$$ \left.\left. {}- \frac{x^2 - 7 x}{12} \,
\mbox{Ei}(-x) \: + \: \frac{x^2 - 14 x - 52}{24} \,
\mbox{Ei}(-x/2)\right\}\right|_{\textstyle x=\frac{p^2}{\Lambda^2}}
\eqno(\mbox{A}.5) $$
\noindent and
$$ \delta f_2^{\mbox{\scriptsize 1--loop}}(p^2) \;\: = \;\: 
\frac{3 g_\Lambda^2}{(4\pi)^2} \, \Bigg\{ \frac{3}{4} \Big(\ln (
x/2) + C \Big) \: - \: \frac{3}{8} $$
$$ {}+ \frac{3 x + 2}{4 x^2} \: + \: \frac{x^2 - x + 2}
{4 x^2} \, e^{-x} \: - \: \frac{x^2 + x + 2}{2 x^2} \, e^{-x/2}
 $$
$$ {}+ \frac{x}{4} \, \mbox{Ei}(-x) \: - \:
\frac{x + 3}{4} \, \mbox{Ei}(-x/2) \Bigg\}
\Bigg|_{\textstyle x=\frac{p^2}{\Lambda^2}} \ \ \ , \eqno(\mbox{A}.6) $$
\noindent where $C$ denotes Euler's constant, $C \cong 0.577216$. The
exponential integral function Ei is defined by
$$ \mbox{Ei}(-x) = \int\limits_{-\infty}^{-x} dt \, \frac{e^t}{t}
\eqno(\mbox{A}.7) $$
\noindent for $x > 0$.



\section*{Figure captions}

\newcounter{fig}
\begin{list}{\bf Figure \arabic{fig}:}{\usecounter{fig}}

\item Diagrammatic form of the ERGEs for the ghost propagator function
$f_{2,k}$ and the gluon propagator function $f_{1,k}$. Internal curly
lines denote the full gluon propagator, internal dotted lines the full
ghost propagator and full points the full vertices from the ansatz (3.10)
for $\widehat{\Gamma}_k$. The crossed circles denote insertions of 
$\partial_k R_{k,\mu \nu}$ resp. $\partial_k \tilde R_k$. 

\item Results for the k--dependent function $F_k (p^2)$ (5.1), which is
related to the potential $V_k (p^2)$ through (5.2), for different values
of $k^2$, $k^2 = \Lambda^2$, 
$10^{-1} \Lambda^2$, $10^{-2} \Lambda^2$, $10^{-3} \Lambda^2$,
$k^2 = {k_0}^2 = 7.43 \cdot 10^{-5} \Lambda^2$, as full lines, and for
$k^2 = \hat{k}^2 = 3.68 \cdot 10^{-5} \Lambda^2$ as a dashed line. The bare 
coupling $g_{\Lambda}$ is $g_{\Lambda} = 1.4$, and $p^2$ is
given in units of $\Lambda^2$. The
curves with decreasing values at $p^2 = 0$ correspond to decreasing
values of $k^2$.

\item As in fig.\ 2, with a maximal value of $p^2 = 10^{-1} \Lambda^2$. 

\item As in fig.\ 2, with a maximal value of $p^2 = 10^{-2} \Lambda^2$ 
in order to resolve the small $p^2$ region.

\item Fit of our result for $F_{k_0} (p^2)$ (full line) to the 
parametrization of Richardson [24] (dashed line), with an optimized
relation between $\Lambda$ and $\Lambda_R \cong 400$ MeV such that
$\Lambda = 12.1$ GeV.   

\item As in fig.\ 5, with a maximal value of $p^2 = 4$ GeV$^2$.

\item  Results for $F_{k_0}(p^2)$ for different values of the bare
coupling  $g_{\Lambda} = 1.4$ (full line), 
$g_{\Lambda} = 1.6$ (long--dashed line), $g_{\Lambda} = 1.8$
(short--dashed line) and $g_{\Lambda} = 2.0$ (dotted line). In each
case we determined the physical value of the starting scale $\Lambda$ by
a fit to the parametrization of Richardson, eq.\ (5.4).

\item As in fig.\ 7, but without the one--loop improvement of the
propagator functions $f_{i,k}$ at the starting scale $\Lambda$. 

\item Results for the propagator functions $f_{1,k_0}$ (full line)
and $f_{2,k_0}$ (dashed line) for $g_{\Lambda} = 1.4$. 

\end{list}

\begin{figure}[p]
\unitlength1cm
\begin{picture}(15,7.5)
\put(-1.4,6.2){$\partial_k$}
\put(-1.4,2.7){$\partial_k$}
\put(2,6.2){$=$}
\put(2,2.7){$=$}
\put(9.1,2.7){$-\qquad\frac{1}{2}$}
\put(3.8,.6){$-$}
\put(9.1,.6){$-$}
\put(9.1,6.2){$+$}
\put(-1.6,-10.2){
\epsfxsize=22cm
\epsfysize=14cm
\epsffile{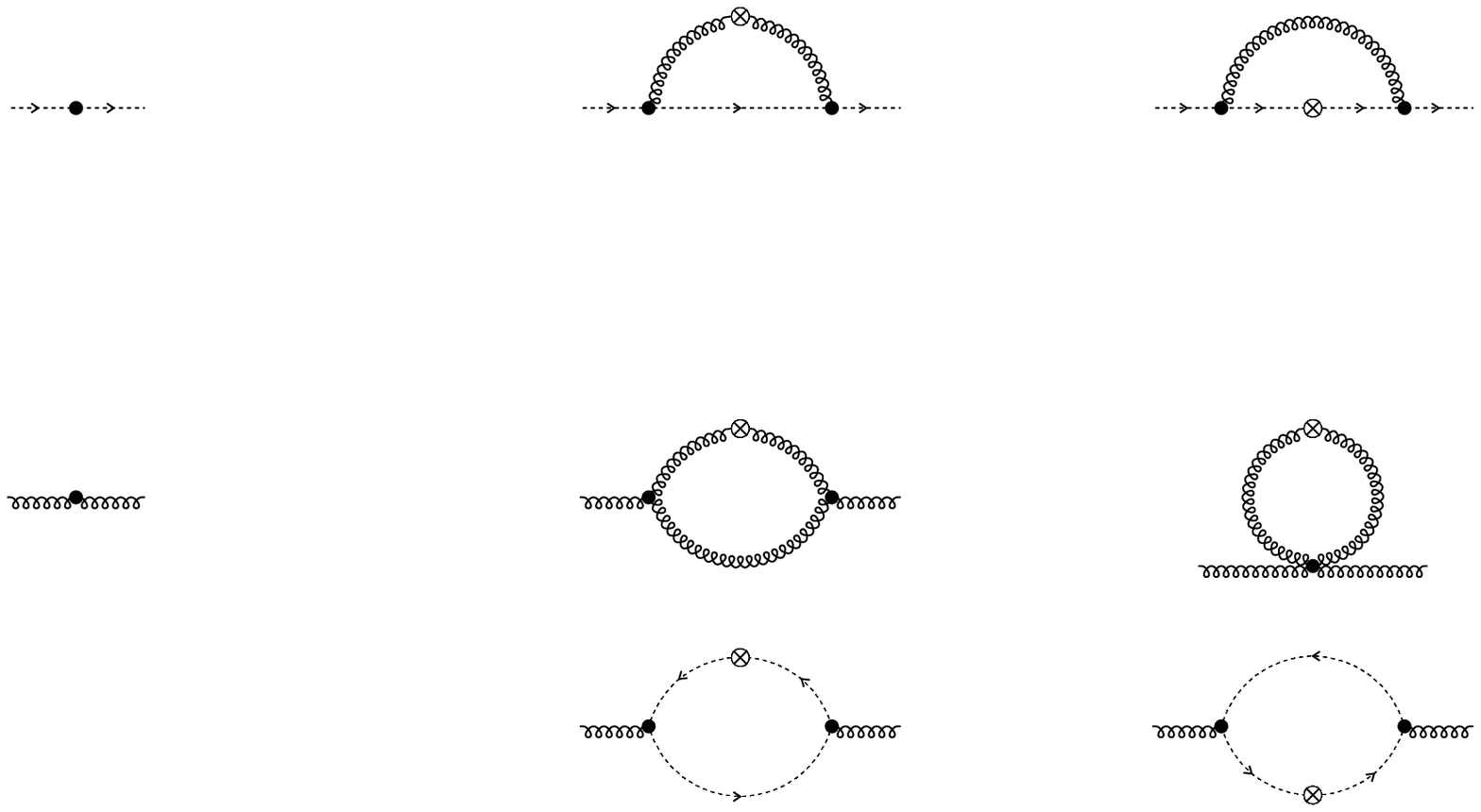}}
\end{picture}\par

\begin{center} \bf Figure 1 \end{center}
\end{figure}

\begin{figure}[p]
\unitlength1cm
\begin{picture}(15,7.5)
\put(12.6,0){$p^2/{\Lambda ^2}$}
\put(-1.3,7.3){$F_{k}(p^2)$}
\put(-1.7,-4.5){
\epsfxsize=14cm
\epsfysize=16cm
\epsffile{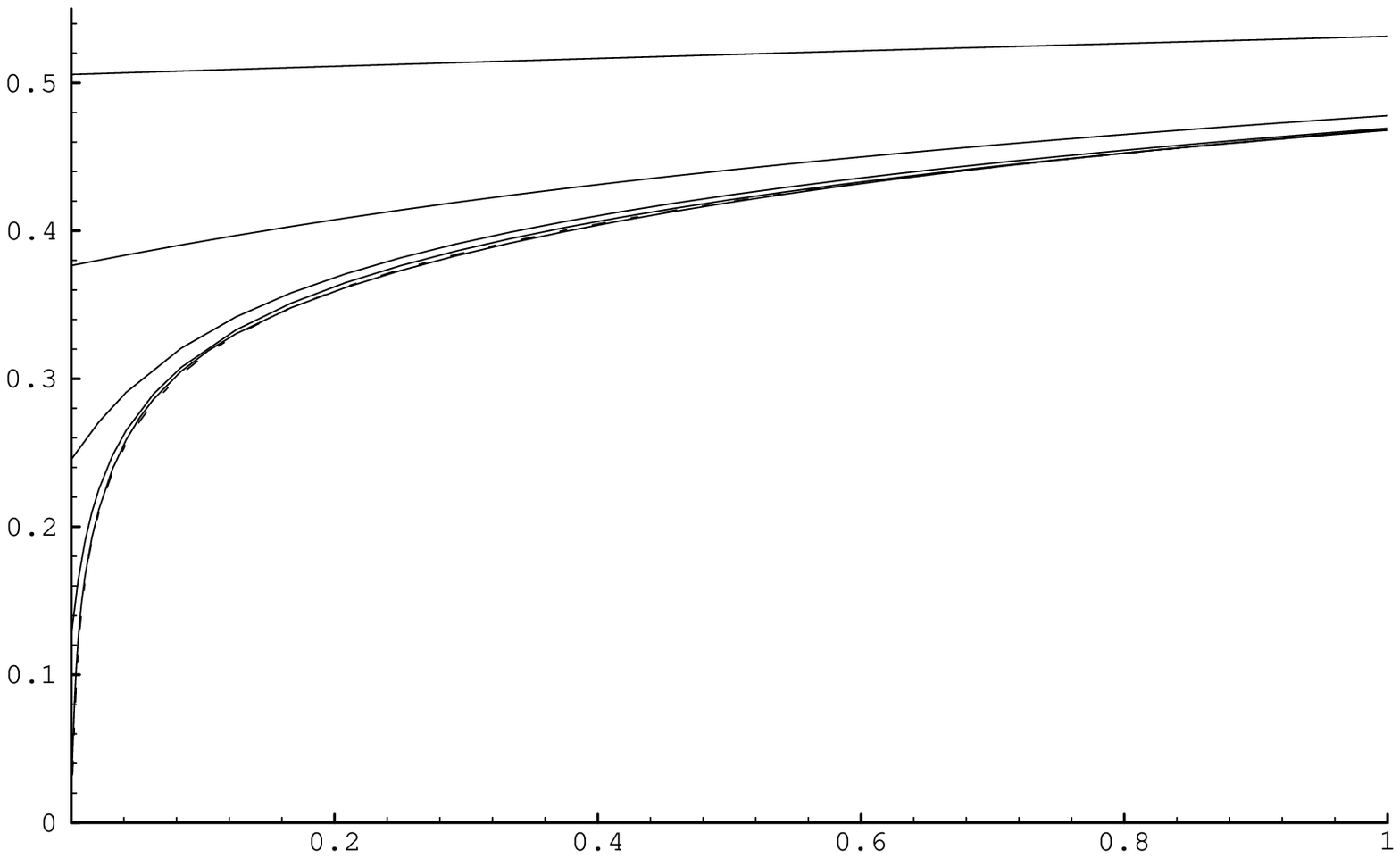}}
\end{picture}\par
\begin{center} \bf Figure 2 \end{center}
\end{figure}

\begin{figure}[p]
\unitlength1cm
\begin{picture}(15,7.5)
\put(12.6,0){$p^2/{\Lambda ^2} $}
\put(-1.3,7.3){$F_{k}(p^2)$}
\put(-1.7,-4.5){
\epsfxsize=14cm
\epsfysize=16cm
\epsffile{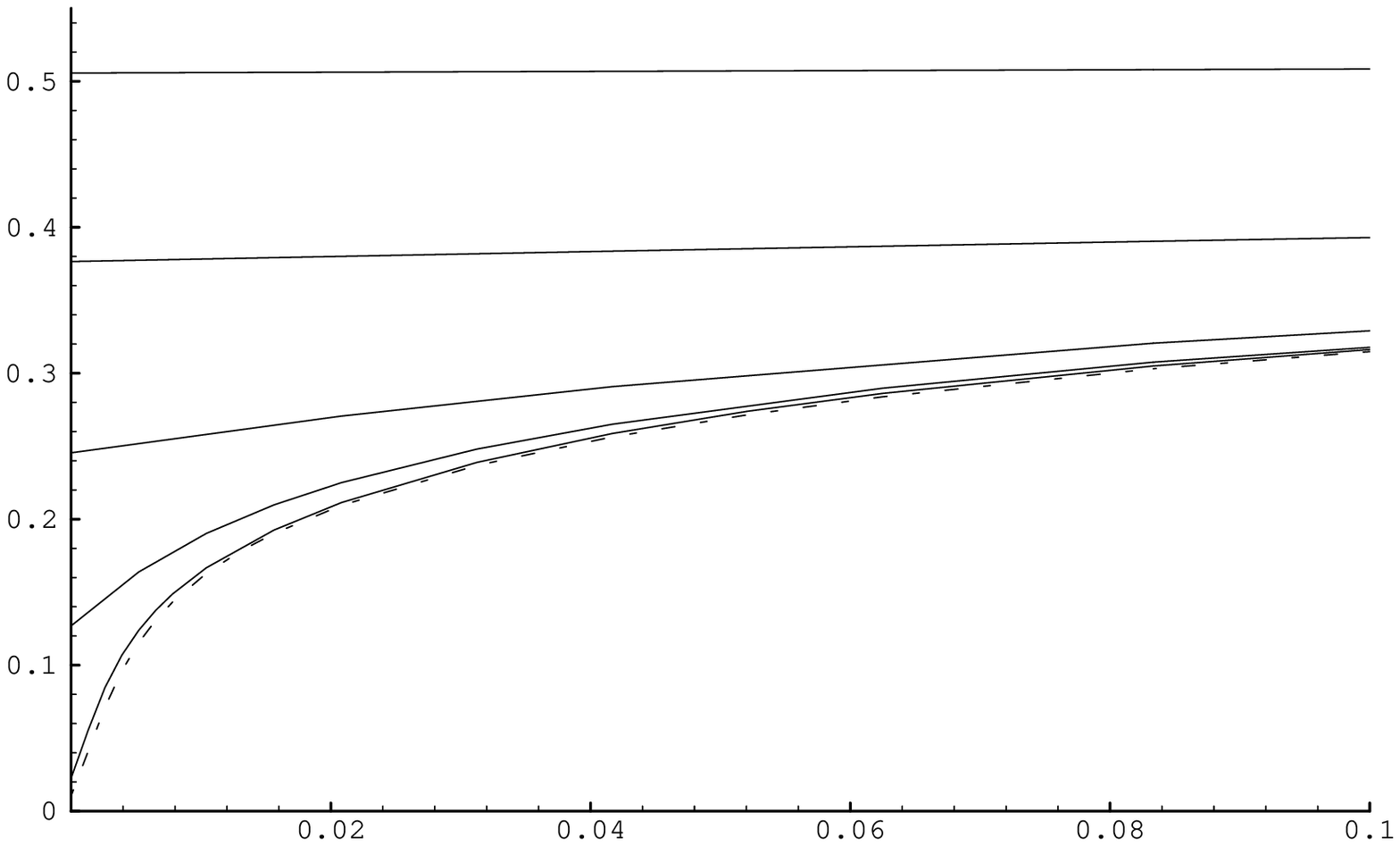}}
\end{picture}\par
\begin{center} \bf Figure 3 \end{center}
\end{figure}

\begin{figure}[p]
\unitlength1cm
\begin{picture}(15,7.5)
\put(12.6,0){$p^2/{\Lambda ^2} $}
\put(-1.3,7.3){$F_{k}(p^2)$}
\put(-1.7,-4.5){
\epsfxsize=14cm
\epsfysize=16cm
\epsffile{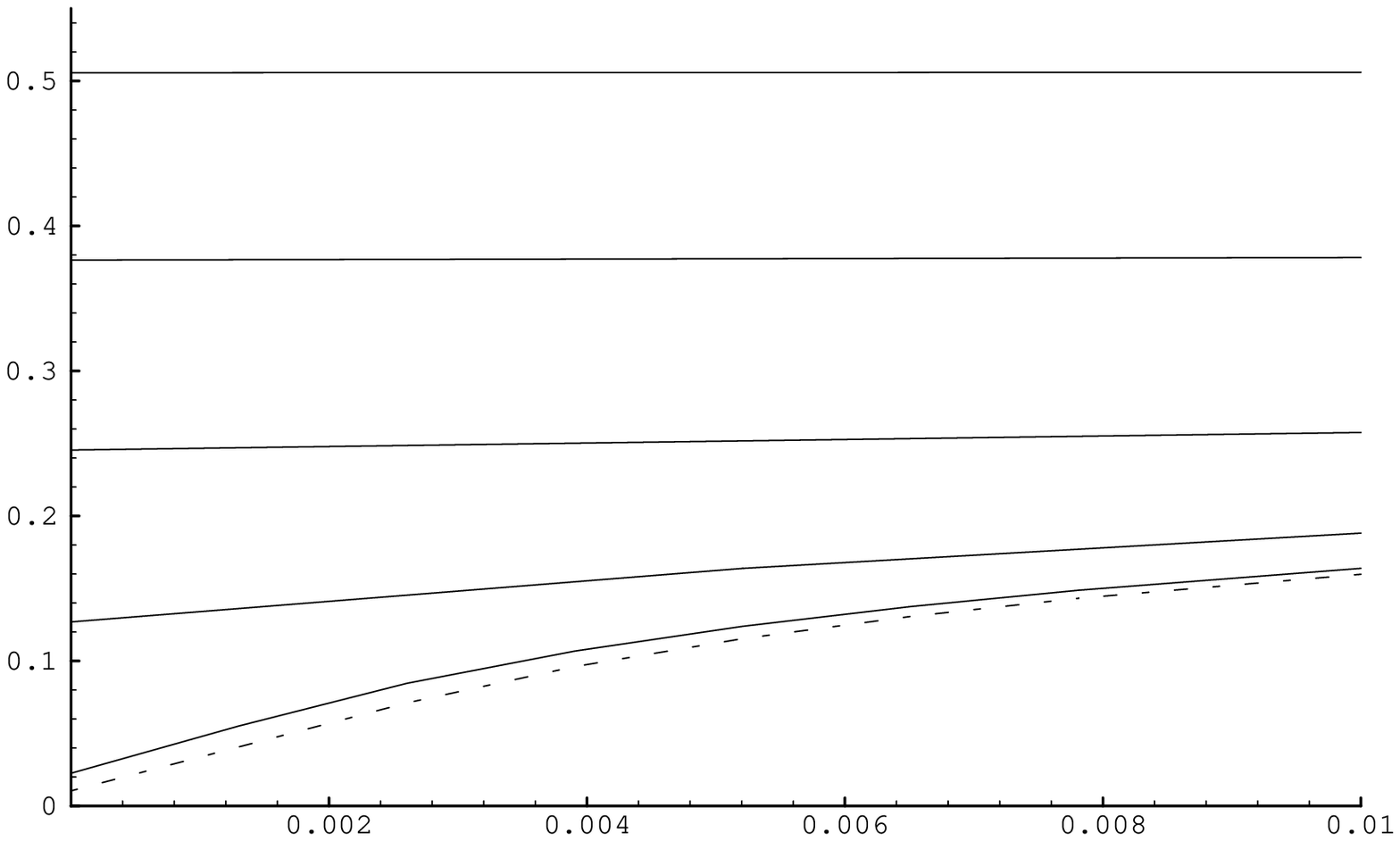}}
\end{picture}\par
\begin{center} \bf Figure 4 \end{center}
\end{figure}

\begin{figure}[p]
\unitlength1cm
\begin{picture}(15,7.5)
\put(12.6,0){$p^2 \;[\mbox{GeV}^2]$}
\put(-1.3,7.3){$F_{k_0}(p^2)$, $F_R(p^2)$}
\put(-1.7,-4.5){
\epsfxsize=14cm
\epsfysize=16cm
\epsffile{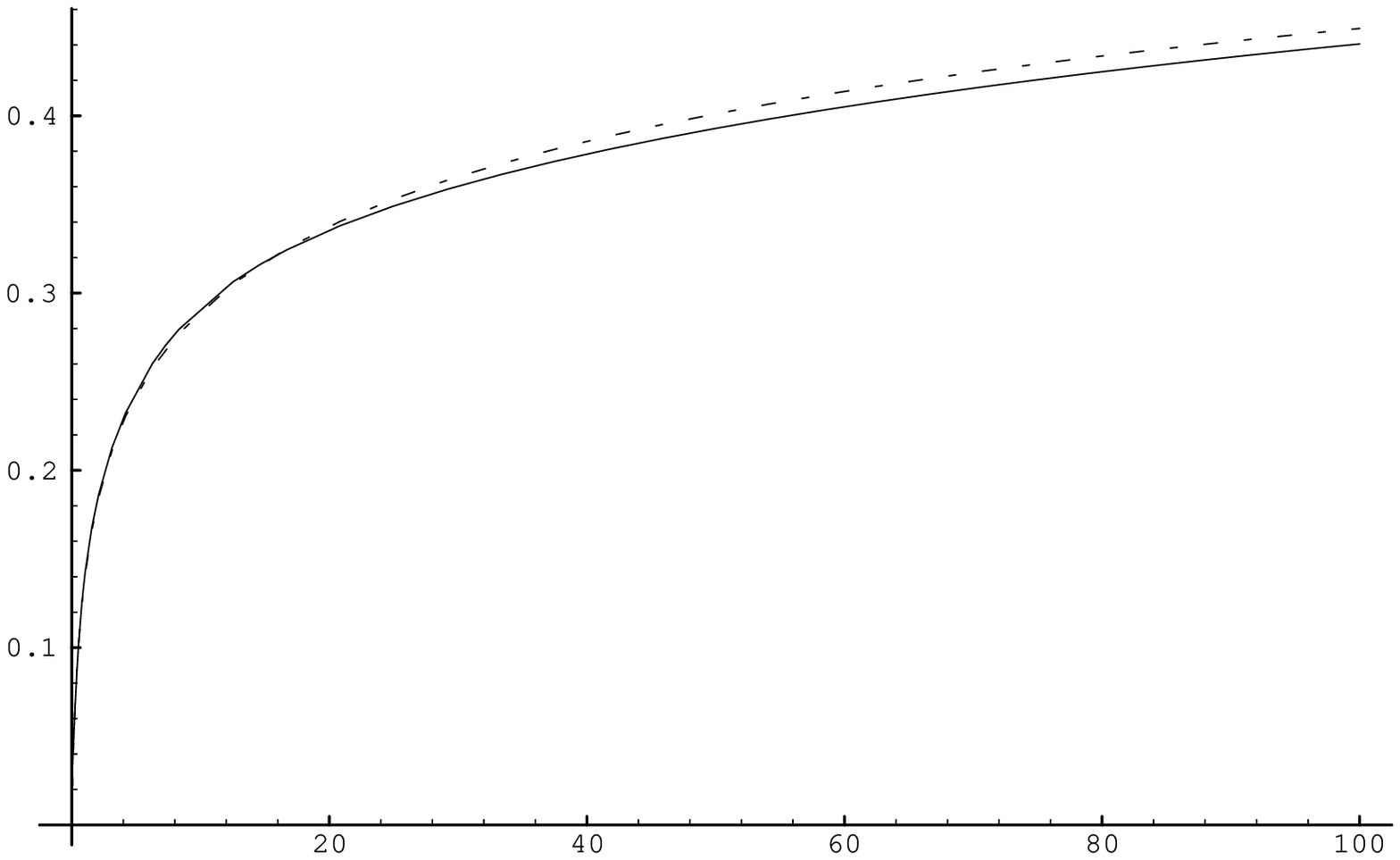}}
\end{picture}\par
\begin{center} \bf Figure 5 \end{center}
\end{figure}

\begin{figure}[p]
\unitlength1cm
\begin{picture}(15,7.5)
\put(12.6,0){$p^2 \;[\mbox{GeV}^2]$}
\put(-1.3,7.3){$F_{k_0}(p^2)$, $F_R(p^2)$}
\put(-1.7,-4.5){
\epsfxsize=14cm
\epsfysize=16cm
\epsffile{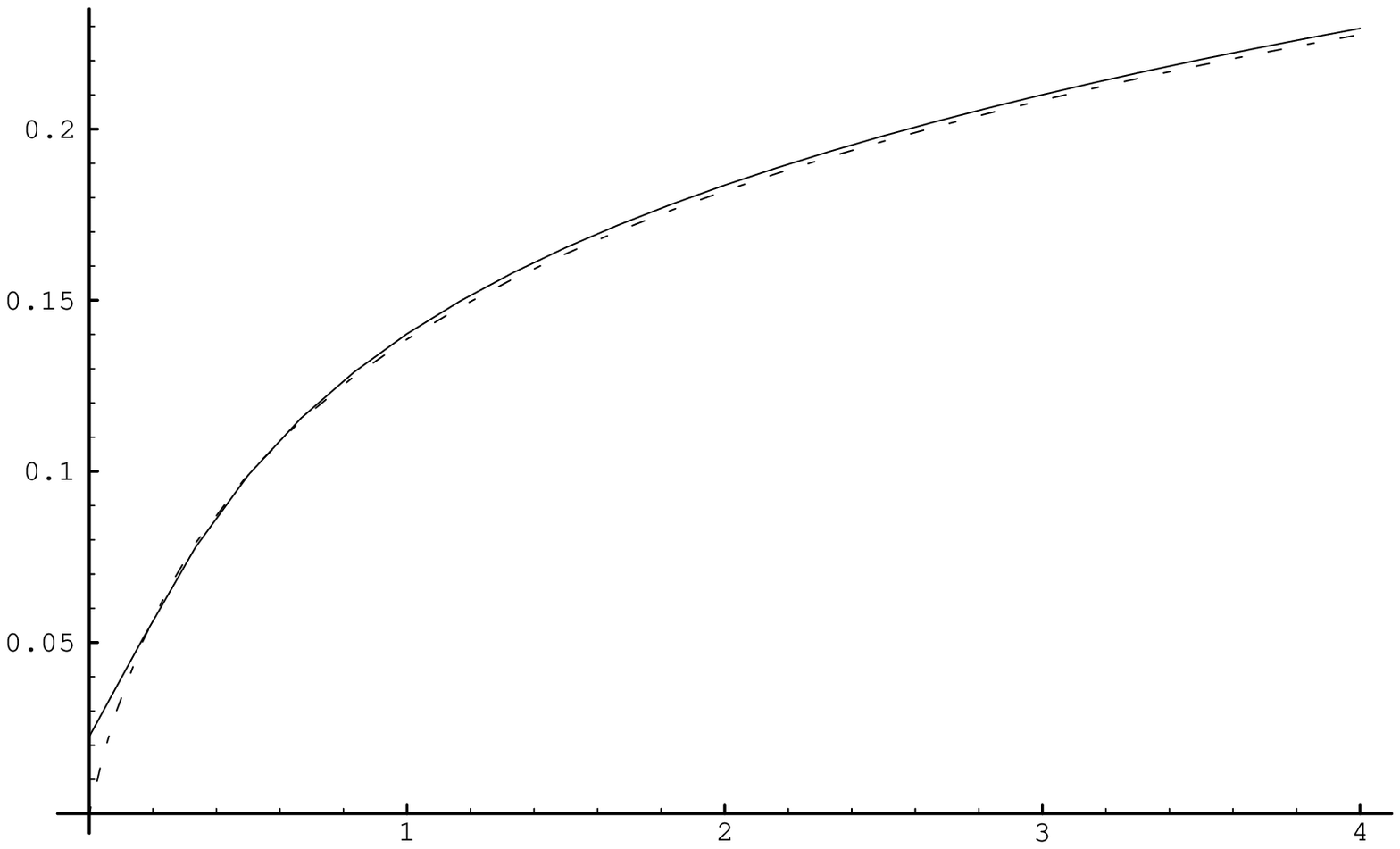}}
\end{picture}\par
\begin{center} \bf Figure 6 \end{center}
\end{figure}

\begin{figure}[p]
\unitlength1cm
\begin{picture}(15,7.5)
\put(12.6,0){$p^2 \;[\mbox{GeV}^2]$}
\put(-1.3,7.3){$F_{k_0}(p^2)$}
\put(-1.7,-4.5){
\epsfxsize=14cm
\epsfysize=16cm
\epsffile{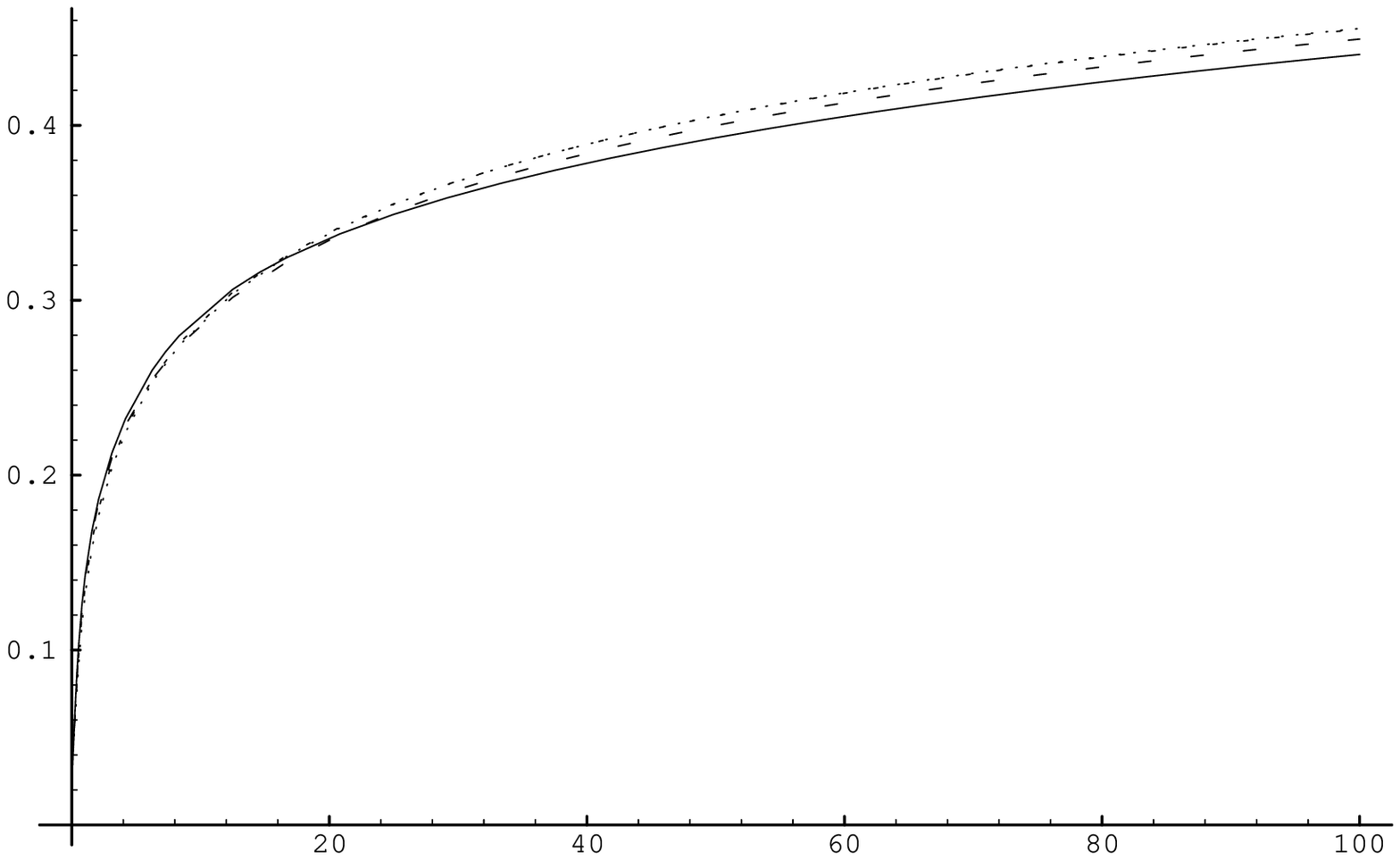}}
\end{picture}\par
\begin{center} \bf Figure 7 \end{center}
\end{figure}

\begin{figure}[p]
\unitlength1cm
\begin{picture}(15,7.5)
\put(12.6,0){$p^2 \;[\mbox{GeV}^2]$}
\put(-1.3,7.3){$F_{k_0}(p^2)$}
\put(-1.7,-4.5){
\epsfxsize=14cm
\epsfysize=16cm
\epsffile{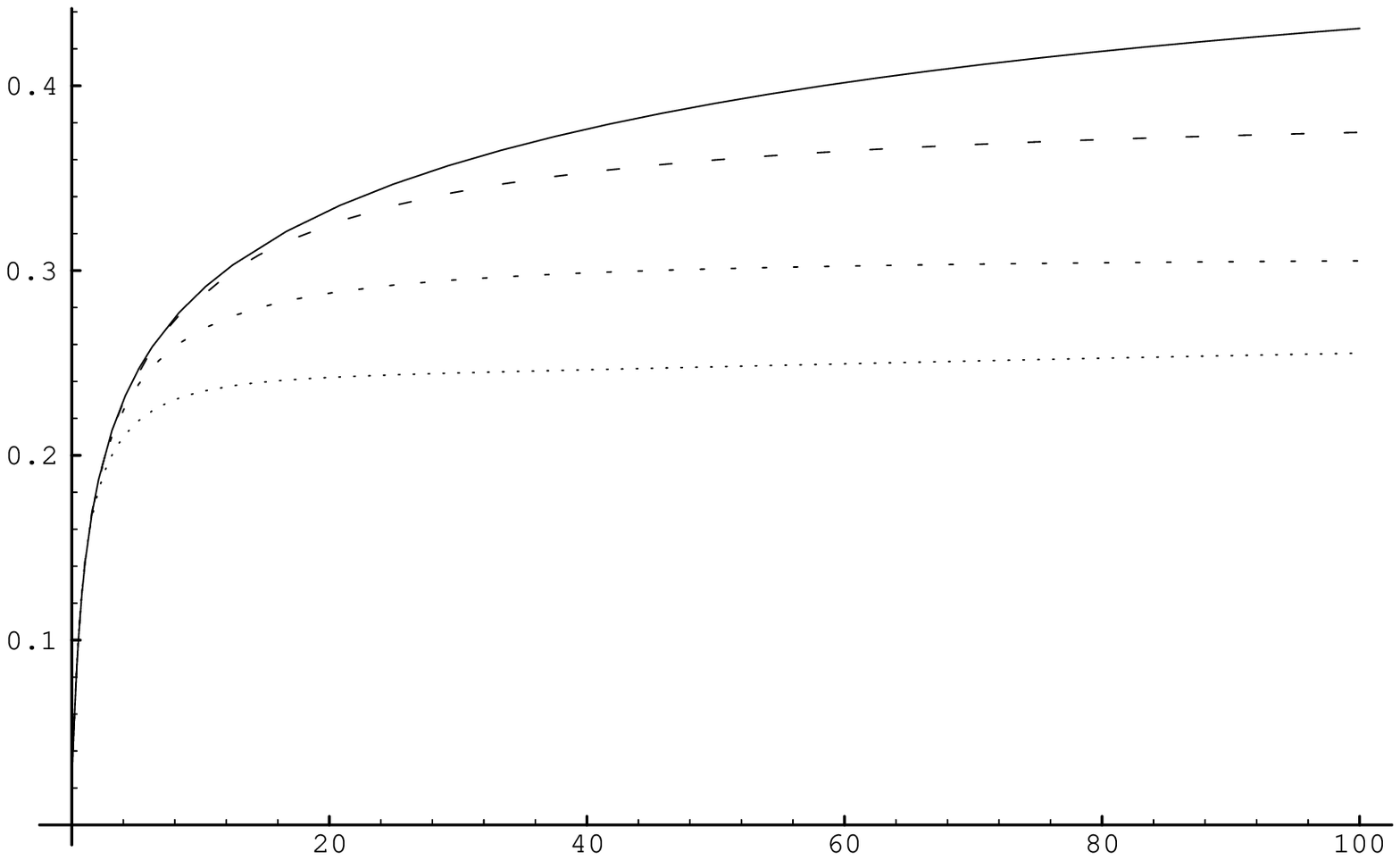}}
\end{picture}\par
\begin{center} \bf Figure 8 \end{center}
\end{figure}

\begin{figure}[t]
\unitlength1cm
\begin{picture}(15,7.5)
\put(12.6,0){$p^2 \;[\mbox{GeV}^2]$}
\put(-1.3,7.3){$f_{1,k_0}(p^2), f_{2,k_0}(p^2)$}
\put(-1.7,-4.5){
\epsfxsize=14cm
\epsfysize=16cm
\epsffile{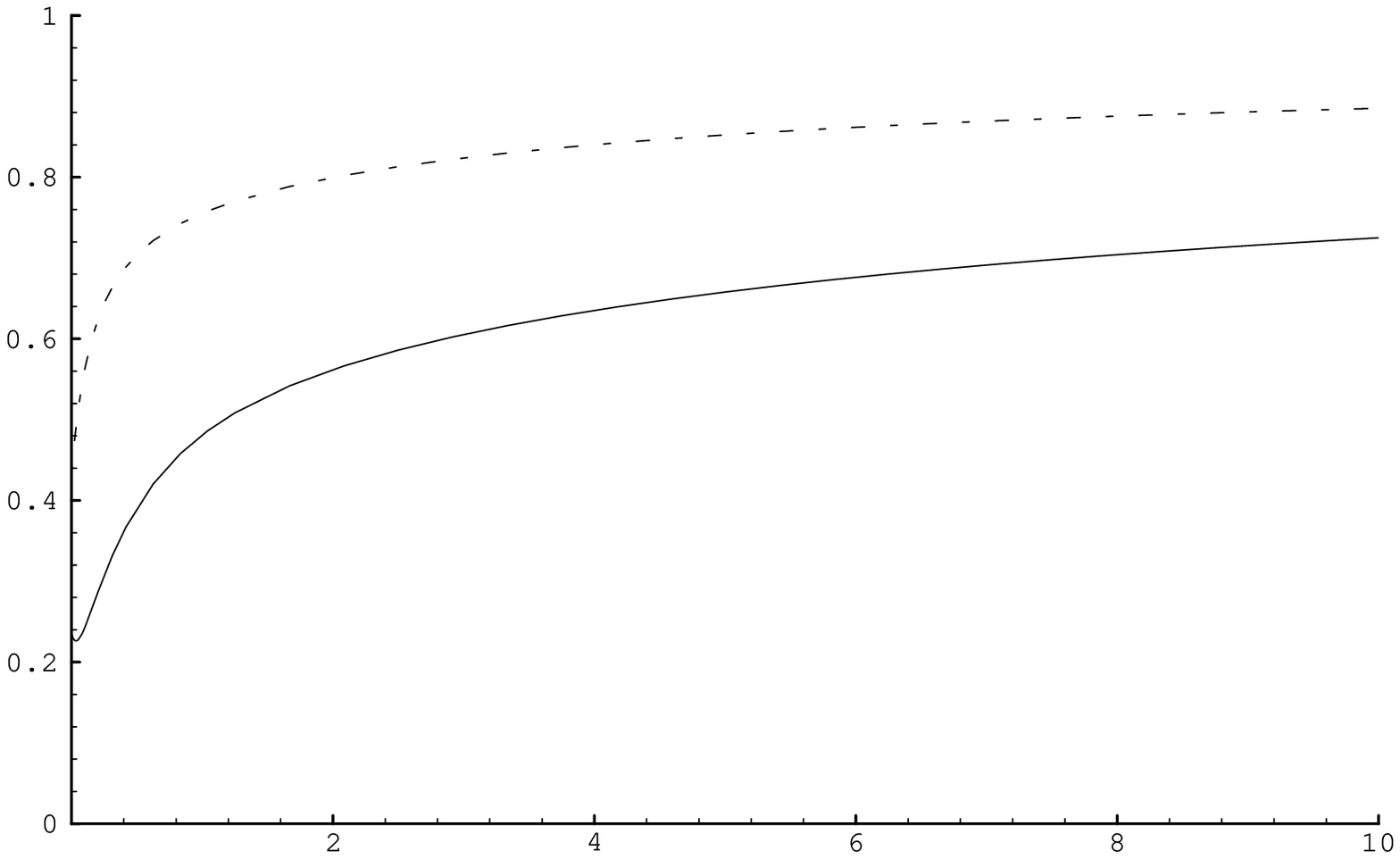}}
\end{picture}\par
\begin{center} \bf Figure 9 \end{center}
\end{figure}

\end{document}